%% file: main.tex
\documentclass[sigconf]{acmart} 
\AtBeginDocument{%
  }

\copyrightyear{2026}
\acmYear{2026}
\setcopyright{cc}
\setcctype{by}
\acmConference[CCS '26]{Proceedings of the 2026 ACM SIGSAC Conference on Computer and Communications Security}{November 15--19, 2026}{The Hague, Netherlands}
\acmBooktitle{Proceedings of the 2026 ACM SIGSAC Conference on Computer and Communications Security (CCS '26), November 15--19, 2026, The Hague, Netherlands}
\acmDOI{10.1145/3830454.3846524}
\acmISBN{979-8-4007-2871-6/2026/11}



\input{packages}
\input{notions}

\fullversion

\begin{document}


\title{Privacy Leakage from a Thousand Words: Millipixel Location Recovery from Dot Maps}



\author{Yuntao Du}
\authornote{Equal contribution.}
\affiliation{%
    \institution{Purdue University}
    \city{West Lafayette}
    \state{Indiana}
    \country{USA}
}
\email{ytdu@purdue.edu}

\author{Tanishq Pauskar}
\authornotemark[1]
\affiliation{%
    \institution{Purdue University}
    \city{West Lafayette}
    \state{Indiana}
    \country{USA}
}
\email{tpauskar@purdue.edu}

\author{Hao Wang}
\affiliation{%
    \institution{Purdue University}
    \city{West Lafayette}
    \state{Indiana}
    \country{USA}
}
\email{wang5329@purdue.edu}

\author{Jing Su}
\affiliation{%
    \institution{Indiana University School of Medicine}
    \city{Indianapolis}
    \state{Indiana}
    \country{USA}
}
\email{su1@iu.edu}

\author{Ninghui Li}
\affiliation{%
    \institution{Purdue University}
    \city{West Lafayette}
    \state{Indiana}
    \country{USA}
}
\email{ninghui@purdue.edu}


\input{0_abs}

\maketitle

\input{1_intro}

\input{2_background}

\input{3_problem}

\input{4_0_method}

\input{5_0_eval}

\input{6_defense}

\input{0_related}
\input{7_discussion}

\begin{acks}
This work was funded in part by the National Science Foundation (NSF) awards CNS-2207204 and CNS-2247794. Any opinions, findings, conclusions, or recommendations expressed in this material are those of the authors and do not necessarily reflect the views of the sponsors.
\end{acks}

\bibliographystyle{ACM-Reference-Format}
\bibliography{ref}

\appendix
\input{0_open}

\input{0_ethic}
\ifcameraready\else
\input{8_0_appendix}
\fi

\end{document}

%% file: packages.tex
\usepackage{xspace,url}

\usepackage{amsmath, amsthm}
\usepackage{subfigure}

\usepackage{caption}
\usepackage{cleveref}

\usepackage[utf8]{inputenc}
\usepackage{algorithm}
\usepackage{algorithmic}
\allowdisplaybreaks

\usepackage{multirow}

\usepackage{enumitem}
\theoremstyle{plain}

\setlist[itemize]{leftmargin=*,noitemsep, topsep=0pt}

\setlist[enumerate]{leftmargin=*,noitemsep, topsep=0pt}

%% file: notions.tex
\newcommand{\ie}{\emph{i.e., }}
\newcommand{\eg}{\emph{e.g., }}

\newcommand{\argmin}{\mathop{\mathrm{argmin}}}

 \newcommand{\stdv}[1]{{\scriptsize$\pm$#1}}

\newcommand{\mypara}[1]{\smallskip\noindent\textbf{#1.} \xspace}

\newcommand{\myquestion}[1]{\smallskip\noindent\textbf{#1?} \xspace}

\newcommand{\mymethod}{\ensuremath{\mathsf{AutoLocate}}\xspace}
\newcommand{\mymethodBK}{\ensuremath{\mathsf{AutoLocate}_{\mathsf{BK}}}\xspace}
\newcommand{\mymethodBU}{\ensuremath{\mathsf{AutoLocate}_{\mathsf{BU}}}\xspace}

\newcommand{\algcomment}[1]{\hfill {\color{blue} $\triangleright$ \emph{\small{#1}}}}

\newcommand{\myfullcomment}[1]{\STATE {\textcolor{gray}{\small\textit{\# #1}}}}

\newcommand{\fullversion}{\global\let\ifcameraready\iffalse}
\newcommand{\cameraversion}{\global\let\ifcameraready\iftrue}

\newcommand{\fullorcamera}[2]{\ifcameraready#2\else#1\fi}
\newcommand{\cameraonly}[1]{\ifcameraready#1\fi}



%% file: 0_abs.tex
\begin{abstract}
Dot maps, which visualize individual data points as dots over a geographic region, are widely used across diverse domains to represent spatial patterns in sensitive data.
However, the understanding of the privacy risks associated with dot maps remains limited, particularly for maps covering large geographic areas.
In this paper, we systematically analyze these risks and present \mymethod, an automated framework for high-precision location recovery.
At its core, \mymethod exploits \textit{anti-aliasing artifacts} introduced during map rendering, which inadvertently encode sub-pixel information about dot locations.
\mymethod formulates location recovery as a black-box optimization problem, iteratively refining estimated coordinates by minimizing perceptual discrepancies over these artifacts between the target map and rendered candidate maps.
Extensive experiments on both real-world and synthetic datasets, across different attack scenarios and a broad range of map configurations (\eg map scale, background, resolution), demonstrate the effectiveness of \mymethod.
In particular, it achieves average recovery errors as low as 1 meter (approximately 0.0002 pixel precision) on small-scale maps of the United States, over $\mathbf{200\times}$ more accurate than existing approaches.
We also propose mitigation strategies and introduce a privacy risk assessment tool to help practitioners evaluate and reduce privacy leakage when publishing dot maps.
\end{abstract}

%% file: 1_intro.tex
\section{Introduction}

A well-known adage in communication is ``A picture is worth a thousand words''.  
A commonly used type of picture is the dot map (also known as the dot distribution/density map), which employs point symbols to visualize the geographic distribution of a large number of related phenomena. 
Dot maps rely on visual scatter to show spatial patterns, especially variations in density. 
They are often used in important fields such as medical research, urban planning, and environmental studies~\cite{chandran2024gis, clark2025geospatial}. 
By representing each instance's location as a ``dot'' (which may take the form of a circle, triangle, or other symbols) on a map, researchers can detect spatial patterns, identify clusters, and trace potential sources of outbreaks~\cite{dotmaplist, soetens2017dot, martinez1989geographic}.
For instance, Soetens et al.~\cite{soetens2017dot} demonstrate the use of dot maps in Germany and the Netherlands by plotting individual disease cases to reveal their spatial distribution and highlight outbreak clusters.
The rapid development of map visualization platforms, ranging from professional Geographic Information Systems (GIS) (\eg ArcGIS~\cite{arcgis} and QGIS~\cite{qgis}) to commercial visualization tools (\eg Tableau~\cite{tableau}) and programming libraries (\eg GeoPandas~\cite{geopandas} and R~\cite{r}), has made it easy to obtain precise geolocation information and publish highly accurate dot maps.

While visualizations from dot maps offer clear insights into spatial relationships, they also raise significant concerns about the privacy of individuals represented on the map. 
This issue becomes particularly critical in privacy-sensitive domains such as healthcare, where dot maps are used for disease surveillance, risk assessment, and monitoring of public health trends~\cite{murad2020gis, chandran2024gis}.
As highlighted in previous studies~\cite{kounadi2014geoprivacy}, the publication of raw geospatial data can introduce serious risks, including threats to personal safety from targeted crimes, legal and ethical violations due to privacy breaches, and social consequences such as neighborhood stigmatization.
Despite these risks, dot maps remain a widely adopted and indispensable spatial visualization and analysis tool, with their use continuing to grow across a broad range of disciplines (see~\Cref{sec:background} for a detailed overview).
Therefore, to balance individual privacy and utility, it is essential to develop methods that can accurately assess privacy risks when publishing dot maps.

Prior studies have investigated these risks by examining how accurately locations can be recovered from dot maps~\cite{kounadi2014geoprivacy, curtis2006spatial, zandbergen2014confidentiality}.
In these works, researchers first identify each dot's centroid using methods such as manual visual inspection~\cite{brownstein2005reverse, curtis2006spatial, leitner2007novices, leitner2006framework} or unsupervised learning~\cite{brownstein2006unsupervised}.
They then encode the estimated centroid to its corresponding geographic coordinates and use the resulting recovery error as a measure of privacy risk.
Using these approaches, several studies~\cite{curtis2006privacy, leitner2006framework} have demonstrated that it is possible to recover individual locations from large-scale dot maps (\ie covering a limited area), with average errors around $100$ meters.
For instance, one study~\cite{leitner2007novices} re-identified residential locations with an average error of $96.38$ meters from a dot map of a parish in the United States.

However, these methods fail on dot maps that cover broad geographic regions.
As shown in our experiments, applying them to maps spanning countries or continents yields recovery errors on the order of hundreds of meters.
To the best of our knowledge, little progress has been made on location recovery from dot maps over the past two decades.
As a result, practitioners may implicitly assume that publishing maps over large areas poses limited privacy concerns.
A concrete example is the U.S. Centers for Disease Control and Prevention (CDC) cartographic guideline~\cite{cdc2012cartographic}, which treats privacy risks differently based on geographic scope.  
For maps depicting small areas, the guidelines recommend omitting locational references such as streets and landmarks, noting that ``confidentiality is more likely to be ensured''. 
In contrast, for maps covering large areas, the guidelines assume that individual points become visually ``imperceptible'', thereby permitting only a general view of geographic distribution ``without enabling identification of an individual''.

In this paper, we propose a powerful location recovery attack that remains accurate \textit{even on dot maps with broad geographic coverage}.
Our key insight is that recovery accuracy can be dramatically improved by exploiting the \textit{anti-aliasing artifacts} produced when rendering dot symbols.
Anti-aliasing~\cite{leler1980human,freeman1974computer} is a standard rendering technique that smooths jagged edges by blending the colors of boundary pixels according to the fraction of each pixel covered by the underlying shape (illustrated in~\Cref{fig:demo}).
While this improves visual quality, the blended colors inadvertently encode sub-pixel information about a dot's geometric centroid, which corresponds to the exact location of the individual represented by the dot.
By reverse-engineering these artifacts, we can recover dot coordinates with precision far surpassing that of previous approaches.

Building on this insight, we introduce \mymethod, an automated location recovery framework that leverages anti-aliasing artifacts to infer precise geographic coordinates from dot maps.
At the core of \mymethod is \textit{perceptual coordinate descent}, an iterative optimization algorithm that refines location estimates by generating candidate maps to minimize the difference in anti-aliasing artifacts between the generated and target dots.
Extensive experiments on synthetic and real population data using three popular visualization tools (\ie QGIS, GeoPandas, and R) demonstrate the effectiveness and robustness of \mymethod across a wide range of map configurations (\eg scales, backgrounds, resolutions, and formats).
In particular, our experiments show that \mymethod dramatically improves location recovery accuracy, achieving errors of \textbf{approximately 1 meter (0.0002 pixel precision)} on maps covering large regions (\eg maps of the United States), outperforming existing methods by up to $\mathbf{200\times}$ in recovery accuracy.

Our work challenges the common assumption in geographic data visualization that scale alone protects privacy, and it highlights the importance of examining how maps are constructed when assessing their privacy risks.
We also explore several mitigation strategies and develop a privacy risk assessment tool.
This tool uses population density information to recommend a coordinate quantization level that meets a target anonymity level, helping practitioners mitigate privacy risks when publishing dot maps.
In summary, we make the following contributions:

\begin{itemize}
    \item We systematically study the privacy risks of dot maps by proposing an automated location recovery framework named \mymethod.
    \item We design a new location recovery algorithm that exploits anti-aliasing artifacts in dot maps for high-precision location estimation, without requiring any knowledge of the map generation or rendering mechanism used.
    \item Extensive experiments show that \mymethod is highly effective, achieving over $200\times$ lower error than prior approaches at recovering dot locations, and remains robust across different map configurations and attack scenarios.
    \item We present an assessment tool to help practitioners evaluate and mitigate the privacy risks of their maps.
\end{itemize}

\mypara{Roadmap}
The rest of this paper is organized as follows.
\Cref{sec:background} provides background on dot maps and their use.
\Cref{sec:threat_model} defines the threat model and attack scenarios.
We then detail our location recovery framework in~\Cref{sec:method}.
\Cref{sec:exp} presents the experimental results of the proposed attacks.
\Cref{sec:mitigation} discusses mitigation strategies and the proposed privacy risk assessment tools.
Related work is detailed in~\Cref{sec:related}, and the paper concludes in~\Cref{sec:conclusion}.

\cameraonly{The full version of this paper can be found at: \href{https://arxiv.org/xxxxx}{https://arxiv.org/xxxxx}.}

\input{fig/combined_demo}

%% file: fig/combined_demo.tex
\begin{figure}[t]
    \centering
    \begin{minipage}[t]{0.27\columnwidth}
        \vspace{0pt} 
        \centering
        \includegraphics[width=\linewidth]{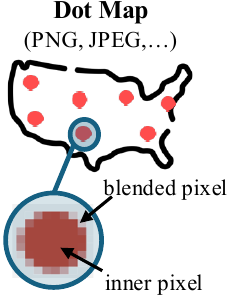}
        \captionof{figure}{Anti-aliasing in dot map.}
        \label{fig:demo}
    \end{minipage}
    \hfill
    \begin{minipage}[t]{0.7\columnwidth}
        \vspace{0pt} 
        \centering
        \captionof{table}{Widespread use of raster dot maps across domains and sensitive attributes.}
        \label{tab:survey}
        \resizebox{\linewidth}{!}{%
        \begin{tabular}{llc}
        \toprule
        \textbf{Application} & \textbf{Sensitive Attribute} & \textbf{Example} \\
        \midrule
        Public Health & Patient home addresses & \cite{buamithup2025lumpy,multidrug-resistant,foraker2022enabling,obaldia2015panama} \\
        Criminology & Crime incident locations & \cite{kebede2024crime, eck2005mapping} \\
        Ecology & Species habitats & \cite{viljanen2024joint,papecs2007modelling,montalvo2025reproducible} \\
        Social Science & Demographics & \cite{dmowska2019racial,leetaru2013mapping} \\
        Education & Student locations & \cite{agostinelli2024spatial, yuan2020application} \\
        Archaeology & Artifact find spots & \cite{jochim2023dots,christoph23,bilotti2024point} \\
        \bottomrule
        \end{tabular}%
        }
    \end{minipage}
\end{figure}

%% file: 2_background.tex
\section{Background}
\label{sec:background}

\mypara{Widespread Use of Dot Maps}
Dot maps have long served as a fundamental tool for revealing spatial patterns and supporting decision-making across diverse disciplines.
One of the most famous early examples is John Snow’s 1854 Broad Street cholera map~\cite{snow1854cholera}, which plotted individual cholera cases as dots, enabling the visual identification of a contaminated water pump as the source of the outbreak.
While this seminal work laid the foundation for modern epidemiology, the utility of dot maps today extends far beyond medical research; they are now an essential visualization method employed by government agencies, major media outlets, and researchers to communicate complex spatial data.

To demonstrate their prevalence, we surveyed recent publications and reports (see \fullorcamera{Appendix~\ref{appendix:survey}}{the full version} for methodology), with~\Cref{tab:survey} providing examples across these varied domains.
We have several key observations from this survey:
(i) Dot maps are widely used to visualize highly sensitive attributes, such as patient home addresses, crime incident locations, and household demographics.
(ii) Despite the sensitive nature of this location data and the widespread reliance on dot maps, none of the surveyed publications explicitly document data protection practices for these visualizations.
(iii) All published dot maps in~\Cref{tab:survey} are \textit{raster images}, representing the map as a grid of colored pixels. 
This dominance is a practical necessity: when organizations visualize dense populations containing thousands or millions of points, rasterization collapses massive spatial datasets into a single, fixed-resolution image. 
This ensures storage efficiency, platform compatibility, and suitability for print and publication.

The extensive use of raster dot maps for sensitive data highlights the critical need to systematically assess their privacy risks. Consequently, this paper focuses on analyzing the specific vulnerabilities of rasterized dot maps. As shown in~\Cref{sec:exp}, our attack achieves high recovery accuracy on dot maps across various scales and configurations, revealing significant privacy flaws in this widely adopted visualization practice.

\mypara{Anti-Aliasing in Dot Maps}
To render dot shapes on a discrete pixel grid, visualization tools apply spatial anti-aliasing by default~\cite{leler1980human,freeman1974computer}. 
Anti-aliasing is a standard graphics technique designed to smooth the edges of rendered elements, mitigating the jagged artifacts (\ie aliasing) that inherently occur when approximating continuous shapes on a finite-resolution display~\cite{kesten2017evaluating}.
The key idea is to blend the colors of boundary pixels with the underlying map background according to the fraction of each pixel covered by the dot shape~\cite{goral1984modeling,crow1977aliasing}.
This produces intermediate color values along edges, resulting in smoother and more visually natural boundaries. 
Over the past decades, various algorithms have been developed to balance rendering quality and efficiency, such as Supersampling (SSAA)~\cite{ssaa}, Multisampling (MSAA)~\cite{msaa}, and Fast Approximate Anti-Aliasing (FXAA)~\cite{fxaa}.
As a concrete example, SSAA divides each pixel into $n$ sub-samples and determines whether the center of each sub-sample falls inside the dot or on the background. The pixel color is then computed by averaging the colors of these sub-samples:
\begin{equation*}
    \mathbf{I}_{x,y} = \frac{m}{n}\,\mathbf{z} + \left(1 - \frac{m}{n}\right)\mathbf{B}_{x,y},
\end{equation*}
where $\mathbf{z}$ denotes the dot color, $\mathbf{B}_{x,y}$ denotes the background color, and $m$ is the number of sub-samples, out of $n$, whose centers fall inside the dot. 
As the dot moves by a sub-pixel amount, $m$ changes accordingly, so the pixel color encodes the dot's sub-pixel position.

Our key insight is that the subtle color gradients produced by this edge blending can be exploited to infer the locations of dots with greater precision, potentially down to the millipixel level.
While our attack exploits these anti-aliasing artifacts, it does not depend on any specific knowledge of the anti-aliasing algorithms used to generate the target dot map.

%% file: 3_problem.tex
\section{Threat Model and Attack Scenarios}
\label{sec:threat_model}

\mypara{Adversary's Goal}
Given a raster map image $\mathbf{I} \in \mathbb{Z}^{W \times H \times 3}$ (with width $W$, height $H$, and RGB color channels, where each channel contains integer values between 0 and 255), each dot in the image represents the location of an individual. 
The adversary’s objective is to infer the underlying geographic coordinates (\ie latitude and longitude) associated with every dot.

\mypara{Adversary's Capabilities}
We make the following realistic assumptions about the adversary's capabilities:
\begin{itemize}
    \item \textit{Raster Image Access.}
    The adversary has access to the raw pixel data of the raster dot map. In practice, this requires minimal effort: dot maps published on websites or in articles can be saved directly, and maps embedded in PDF documents can be easily extracted at their original resolution using tools like pdfimages~\cite{pdfimages}.
    \item \textit{Dot Properties.} 
    The adversary knows the visual properties of the target dots, including their geometry $\phi$ (\eg circle), size $\rho$, and color $\mathbf{z}$, from which the pixel area of a single dot $\mu$ directly follows.
    These properties are easy to obtain by visual inspection and image editing tools (\eg a pixel selector).
    \item \textit{Map Rendering Tool.}
    The adversary has access to a map visualization tool, modeled as a rendering function $\mathcal{R}$, that is the same as or similar to the tool used to generate the target map. We show in~\Cref{sec:exp_ablation} that the attack remains effective when the adversary's tool differs from the one used to generate the target map. 
    We treat the rendering process as a black box: the adversary needs no knowledge of its internal mechanisms, such as the specific anti-aliasing algorithm.
    \item \textit{Coordinate Transformation.}
    The adversary can learn a coordinate transformation function, \ie $\mathcal{F}:(x, y) \mapsto (\text{lat}, \text{lon})$, which maps coordinates $(x,y)$ in the raster map to geographic coordinates.
    This transformation function can be derived from map legends (which provide scale and projection details) or reconstructed using the georeferencing features of modern GIS tools (\eg QGIS) to align the map with a known coordinate system.
\end{itemize}

Together, these capabilities enable the adversary to render new dot maps and compare them against the target map $\mathbf{I}$ to accurately recover the dot locations, as detailed later in our attacks.

\begin{figure}[t]
  \centering
  \begin{minipage}[]{0.32\linewidth}
    \includegraphics[width=\linewidth]{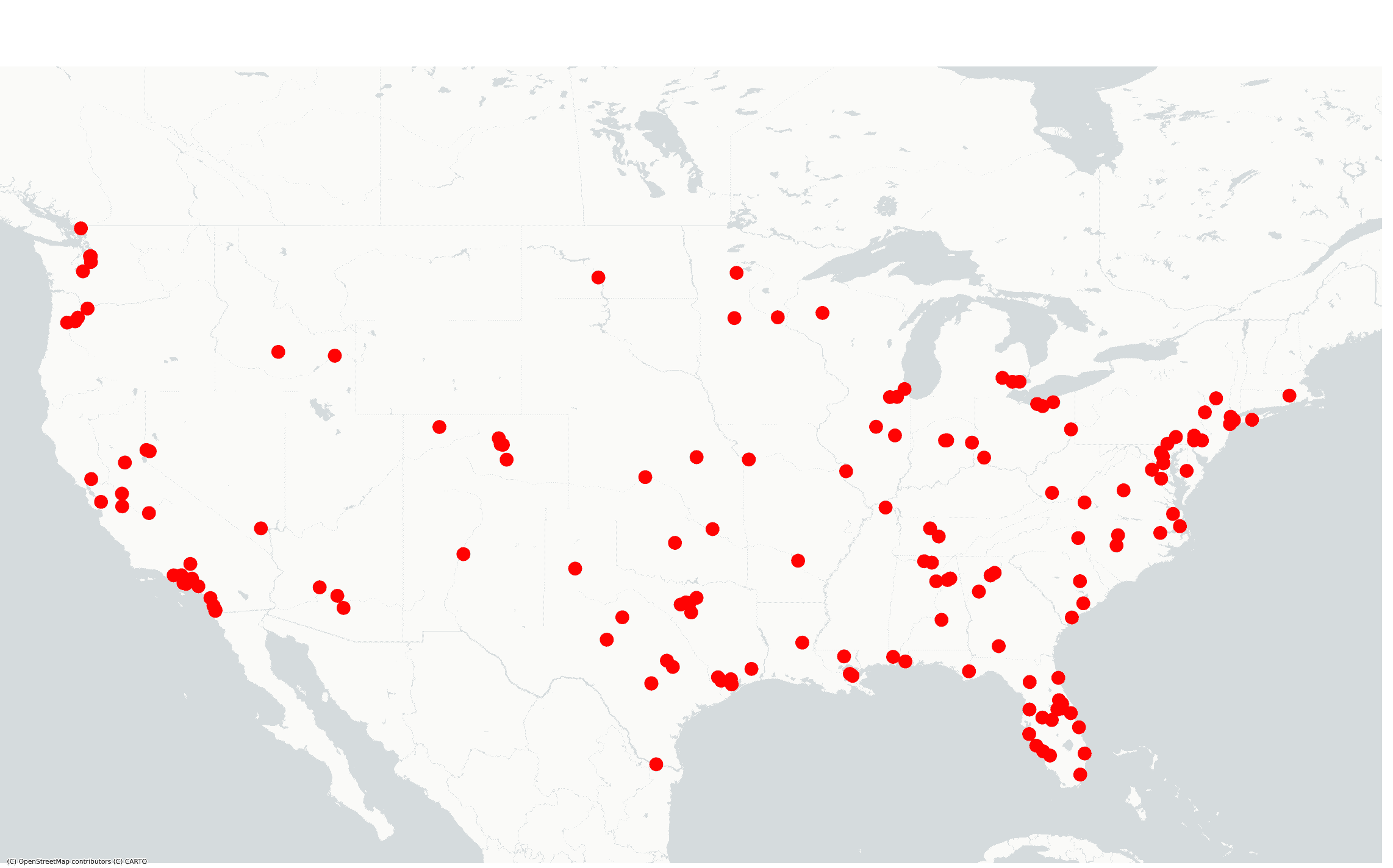}
    \caption*{(a) White canvas}
  \end{minipage}
  \begin{minipage}[]{0.32\linewidth}
    \includegraphics[width=\linewidth]{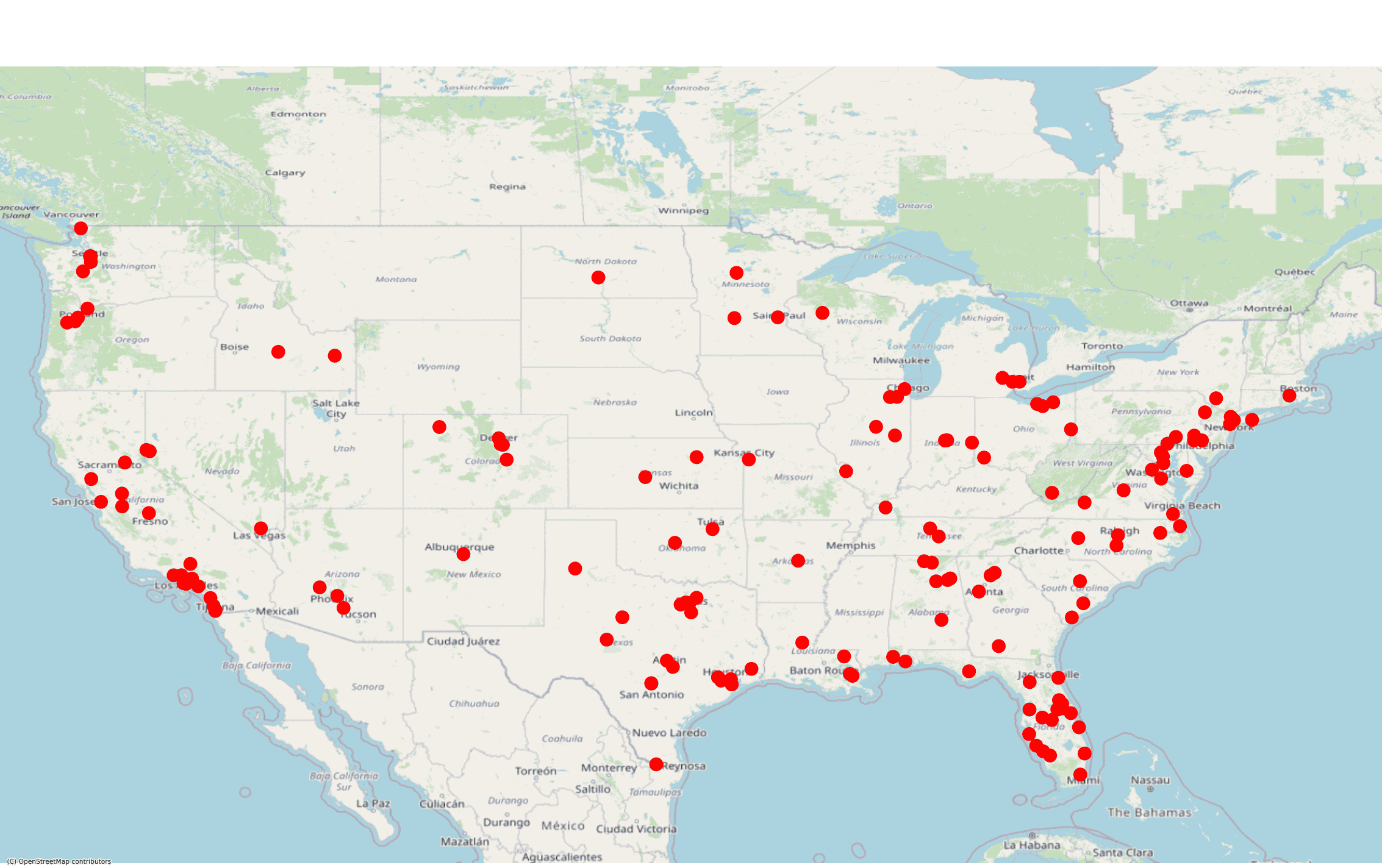}
    \caption*{(b) Street map}
  \end{minipage}
  \begin{minipage}[]{0.32\linewidth}
    \includegraphics[width=\linewidth]{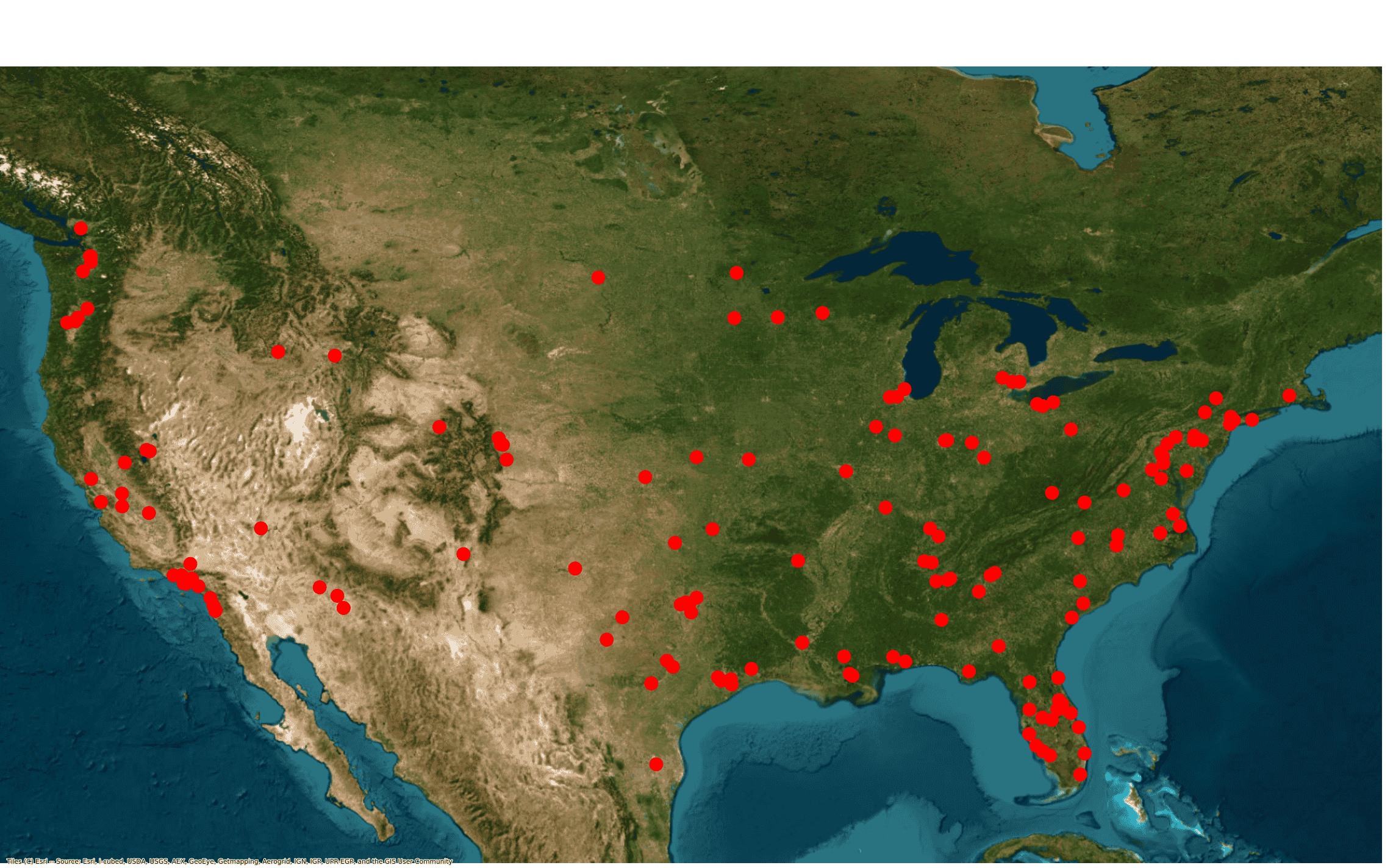}
    \caption*{(c) Satellite map}
  \end{minipage}
  \hfill
  \caption{Examples of three background types of dot maps.}
  \label{fig:map_background}
\end{figure}

\mypara{Attack Scenarios}
As discussed in~\Cref{sec:background}, anti-aliasing blends each dot's boundary pixels with the underlying map background $\mathbf{B} \in \mathbb{Z}^{W \times H \times 3}$, \ie the base map onto which the dots are rendered.
Therefore, knowledge of the background provides
additional signals for recovering dot locations. 
Published dot maps adopt backgrounds of varying complexity, ranging from a blank canvas to a street map or a satellite image, as demonstrated in~\Cref{fig:map_background}.
We distinguish two attack scenarios based on whether the background is available to the adversary:
\begin{itemize}
    \item \textit{Background Known.}
    The adversary has access to the map background $\mathbf{B}$ used to generate the dot map $\mathbf{I}$.
    This is realistic even when the original background is not directly provided: many published dot maps (as shown in~\Cref{tab:survey}) use simple uniform colors (\eg white) or standard basemaps from public repositories (\eg OpenStreetMap~\cite{openstreetmap}) that are easy to replicate.
    \item \textit{Background Unknown.}
    The adversary only has access to the final map $\mathbf{I}$.
    This occurs when the map employs a proprietary or custom background that is not publicly available.
    In this case, the adversary must infer dot locations solely from the target map.
\end{itemize}

Although some map exports include auxiliary data (\eg metadata in TIFF), our attack does not rely on such information.
This ensures that our approach remains effective across common image formats, as shown in~\Cref{sec:exp_main}.


\mypara{Graph Representation of Raster Maps}
We model the raster map $\mathbf{I}$ as a grid graph in which each pixel corresponds to a node, and edges connect each node to its eight spatial neighbors (\ie the horizontally, vertically, and diagonally adjacent pixels).
A \textit{dot region} is then defined as a connected component of the subgraph induced by the pixels whose color matches the dot color $\mathbf{z}$, which can be efficiently identified using standard graph traversal algorithms (\eg breadth-first search).
This formulation provides a notion of pixel adjacency that underpins the operations of our attack.

\input{tables/alg_baseline}

\begin{figure*}[t]
    \centering
    \includegraphics[width=0.98\textwidth]{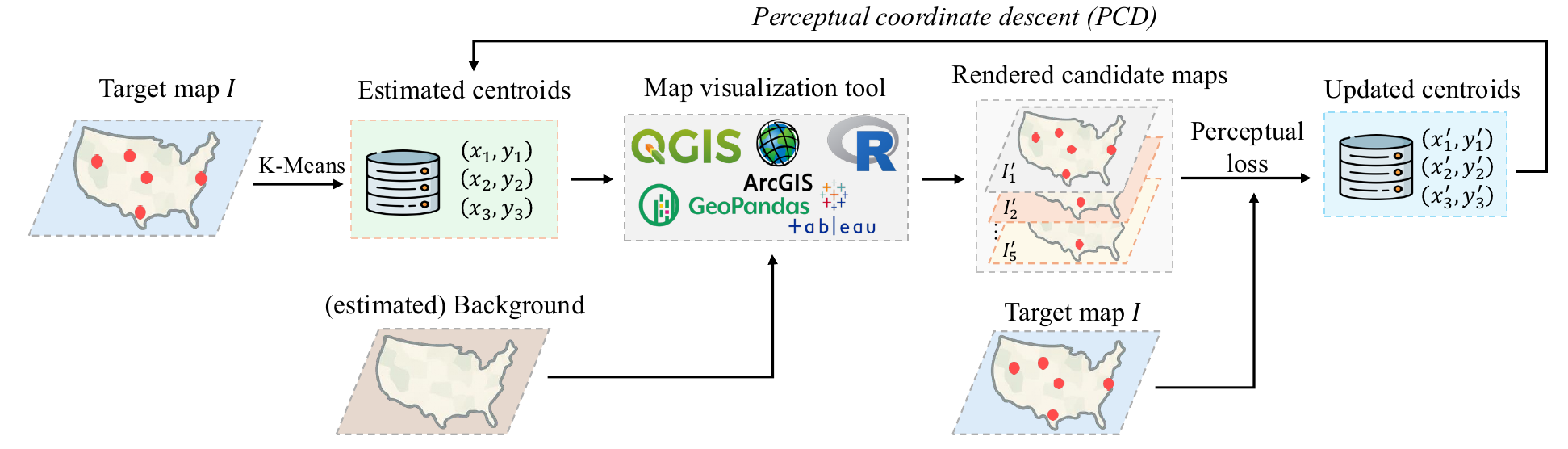}
    \caption{Illustration of the key processes in \mymethod. The adversary first initializes dot locations using K-Means clustering.  Using map visualization tools and the (estimated) map background, the adversary renders candidate maps for five search directions (\ie stay, left, right, up, and
    down). The direction that minimizes the perceptual loss is selected for each dot. This process is repeated to refine the location estimates. Best viewed in color.}
    \label{fig:pba}
\end{figure*}

\mypara{Baseline: PixelMatch}
A straightforward method for estimating dot locations is to compute the centroid of each dot region, which we call PixelMatch. 
As detailed in~\Cref{alg:baseline}, the algorithm first extracts all connected components composed of pixels matching the target color $\mathbf{z}$ (Line 1). 
Each connected component $P$ corresponds to a candidate dot, and the algorithm computes the mean coordinate of its constituent pixels to determine its centroid (Line 4). 
Finally, these centroids are mapped to geographic coordinates using the transformation $\mathcal{F}$ to obtain the estimated locations.

\mypara{Missed Opportunities of Existing Approaches}
Prior studies~\cite{brownstein2005reverse,brownstein2006unsupervised,leitner2007novices} have used the idea of PixelMatch, either through manual inspection or by using GIS tools for centroid estimation.
However, they fail to fully exploit the available information and do not account for realistic scenarios in which dots may overlap:
\begin{itemize}
    \item \textit{Anti-aliasing Artifacts on Dot Boundaries.}
    PixelMatch estimates a dot's location using only its inner pixels, \ie the pixels whose color exactly matches the dot color.
    The set of inner pixels changes only when the dot moves far enough to fully cover or uncover a pixel.
    A range of sub-pixel dot locations therefore produces the same set of inner pixels, from which PixelMatch would estimate the same dot centroid.
    Anti-aliased boundary pixels provide the additional information needed to distinguish these locations, as their color values are sensitive to the dot's sub-pixel position.
    However, existing approaches overlook these boundary pixels because their color does not exactly match the dot color. 
    Moreover, as demonstrated in~\Cref{sec:exp}, simply incorporating boundary pixels into the centroid computation is still ineffective for high-precision location recovery.
    \item \textit{Overlapping Dots.}
    In dense regions, multiple dots may overlap, forming overlapping dot regions where the boundaries of individual dots merge. 
    PixelMatch treats this merged component as one dot, computing a single centroid and failing to recover the individual dot locations within.
\end{itemize}

%% file: tables/alg_baseline.tex
\begin{algorithm}[t]
\caption{\textbf{Baseline: PixelMatch.} The algorithm identifies
connected components of pixels matching the target dot color and
computes the centroid of each as the estimated location.}
\label{alg:baseline}
\begin{algorithmic}[1]
\REQUIRE Target dot map $\mathbf{I}$, dot color $\mathbf{z}$
\STATE $\mathcal{P} \gets \texttt{FindConnectedComponents}(\mathbf{I}, \mathbf{z})$
\STATE $\mathcal{C} \gets \emptyset$ \algcomment{initialize set of estimated locations}
\FOR {each component $P \in \mathcal{P}$}
    \STATE $(x_c, y_c) \gets \frac{1}{|P|} \sum_{(x, y) \in P} (x, y)$
    \STATE $\mathcal{C} \gets \mathcal{C} \cup \{(x_c, y_c)\}$
\ENDFOR
\RETURN $\mathcal{C}$
\end{algorithmic}
\end{algorithm}

%% file: 4_0_method.tex
\section{\mymethod: A Framework for Automated Location Recovery from Dot Maps}
\label{sec:method}

In this section, we present \mymethod, an automated framework for high-precision location recovery from dot maps. 
\mymethod formulates location recovery as an optimization problem: it iteratively adjusts the estimated dot locations to minimize the perceptual discrepancy between the target map and a rendered candidate map, using anti-aliasing artifacts as the optimization signal. 

\input{4_1_gradient}

%% file: 4_1_gradient.tex
\subsection{Attack Method}

We first consider the attack scenario where the adversary possesses the map background $\mathbf{B}$. 
In this setting, the adversary can render new dot maps from a set of estimated coordinates and compare them with the target map. 
By analyzing the discrepancies at anti-aliased boundaries, the adversary iteratively adjusts the coordinates until the rendered map closely matches the target. 
We first define the different types of pixels in dots, formulate the optimization loss function, discuss the handling of overlapping dots, and present the optimization algorithm and the complete framework.

\mypara{Inner Pixels and Boundary Pixels}
For each dot, we identify its \textit{dot region} as the connected component of pixels matching the dot color $\mathbf{z}$ that contains the dot's estimated location.
The pixels in this connected component are the \textit{inner pixels} of the dot: they are fully covered by the dot color and carry no sub-pixel positional information. 
The \textit{boundary pixels} are pixels that (i)~are four-neighbors (\ie up, down, left, or right) of
at least one inner pixel, and (ii)~do not belong to the connected component, \ie their color differs from $\mathbf{z}$.
The detailed procedure for identifying these pixels is in~\fullorcamera{\Cref{alg:boundary}}{the full version}.
These pixels are anti-aliasing artifacts produced by blending the dot color with the underlying map background, and their color values are sensitive to the dot's sub-pixel position. 
We denote the set of boundary pixels for a dot as $\mathcal{S}$.



\mypara{Perceptual Loss Function}
Given a set of estimated dot locations $\mathcal{C}$, the dot's properties (color $\mathbf{z}$, shape $\phi$, size $\rho$), and the background $\mathbf{B}$, the adversary renders a candidate dot map $\mathbf{I}^\prime$ using the map rendering function $\mathcal{R}$:
\begin{equation*}
    \mathbf{I}^\prime \gets \mathcal{R}(\mathbf{B}, \mathcal{C},
    \mathbf{z}, \phi, \rho).
\end{equation*}
A good location estimate should produce a candidate map that perceptually matches the target map $\mathbf{I}$, particularly at the boundary pixels $\mathcal{S}$ where small positional shifts produce measurable color changes. 
We first define the perceptual discrepancy $d(\cdot,\cdot)$ between two pixels $\mathbf{a}$ and $\mathbf{b}$ as the $L_1$ distance between their RGB channels:
\begin{equation*}
    d\left(\mathbf{a}, \mathbf{b}\right)=
    |\mathbf{a}^r-\mathbf{b}^{r}|+
    |\mathbf{a}^g-\mathbf{b}^{g}|+
    |\mathbf{a}^b-\mathbf{b}^{b}|.
\end{equation*}
A boundary pixel at position $(x, y)$ is produced by anti-aliasing, which blends the dot color $\mathbf{z}$ with the background color $\mathbf{B}_{x,y}$; consequently, its color lies between these two values.  
We define the relative color deviation of the pixel from the background, normalized by the maximum possible deviation $d(\mathbf{z}, \mathbf{B}_{x,y})$:
\begin{equation*}
    \delta_{x,y} = \frac{d(\mathbf{I}_{x,y}, \mathbf{B}_{x,y})}{d(\mathbf{z}, \mathbf{B}_{x,y})}.
\end{equation*}
Using the same approach, we compute the corresponding relative deviation $\delta^\prime_{x,y}$ for the candidate map $\mathbf{I}^\prime$. 
For each dot $\mathbf{c} \in \mathcal{C}$, we calculate the total perceptual loss over its boundary pixels $\mathcal{S}$ by summing the absolute differences between these relative deviations:
\begin{equation*}
    \mathcal{L}(\mathbf{I}, \mathbf{I}^\prime, \mathbf{B}, \mathcal{S}) = \sum_{(x,y)\in \mathcal{S}} \left| \delta_{x,y} - \delta^\prime_{x,y} \right|.
\end{equation*}
This focuses optimization on the boundary where anti-aliasing encodes fine-grained positional information.

\mypara{Handling Overlapping Dots}
The above definitions assume each dot occupies its own connected component.
In dense regions, multiple dots may overlap, causing their connected components to merge into a single component.
This raises two challenges: the number of individual dots within the merged component is unknown, and boundary pixels between overlapping dots may be absorbed as inner pixels.
We address this through a two-stage strategy:
\begin{itemize}
    \item \textit{Estimating the Number of Dots.}
    For each connected component $P$, we estimate the number of overlapping dots $k$ by dividing the total pixel area of the region by the area of a single dot ($\mu$).
    We then apply K-Means clustering~\cite{kmeans} to the pixels in $P$, partitioning it into $k$ groups and yielding $k$ initial dot locations.
    \item \textit{Boundary Pixel Assignment.}
    To identify the relevant boundary pixels for each individual dot, we assign each boundary pixel to its nearest estimated dot centroid.
    This ensures that every boundary pixel belongs exclusively to a single dot, preventing contamination from neighboring overlapping dots. 
\end{itemize}

This procedure enables accurate identification of boundaries for overlapping dots, providing a stable signal for exploiting anti-aliasing artifacts in location estimation.

\input{tables/alg_desent}

\mypara{Perceptual Coordinate Descent (PCD)}
While we can compute the perceptual loss for each dot's current estimated location, applying standard gradient-based optimization (\eg SGD~\cite{nature86sgd,nature15deeplearning}) to minimize this loss is infeasible because the rendering function $\mathcal{R}$ is a black box whose gradients are intractable to compute. 
To address this, we propose Perceptual Coordinate Descent (PCD), a gradient-free algorithm inspired by zeroth-order optimization~\cite{ieee02_zo,book12_zo,icml11_dloptimization}.
Instead of computing gradients, PCD probes neighboring positions on the two-dimensional pixel grid and selects the move that most reduces the loss, using the boundary pixels $\mathcal{S}$ as the optimization signals.
This process repeats until no direction produces further improvement or a maximum number of iterations is reached.

\mypara{Framework Overview}
The complete location recovery framework is outlined in~\Cref{alg:framework} and~\Cref{fig:pba}. 
The algorithm consists of two phases. 
Phase~1 (lines 1--8) initializes dot locations. 
It extracts all connected components of pixels matching the dot color $\mathbf{z}$. 
For each component $P$, it estimates the number of dots $k$ and applies K-Means clustering to establish initial coordinates. 
Phase~2 (lines 9--28) iteratively refines locations using PCD.
In each iteration, the algorithm considers five candidate directions (\ie stay, left, right, up, and down) with step size $\eta$, renders a candidate map for each direction (lines 16--19), and evaluates the perceptual loss at each dot's boundary pixels.
The location yielding the smallest loss is selected for each dot (lines 22--25).
This process is repeated for $T$ iterations to ensure convergence.

Note that the algorithm renders one map per search direction, in which all dots are shifted simultaneously.
We also implement a per-dot alternative, where the loss for each dot is computed by rendering a separate candidate map that shifts only that dot. 
We compare the attack performance and efficiency of these two implementations in \fullorcamera{Appendix~\ref{appendix:implementation}}{the full version}.

\mypara{Attack with Unknown Background}
When the adversary does not possess the map background $\mathbf{B}$, the optimization approach cannot be directly applied. 
To address this, we adapt the framework by estimating the background color of boundary pixels from the target map.
Specifically, we first use the same boundary detection procedure \fullorcamera{(\Cref{alg:boundary})}{described in the full version} to identify the boundary pixels $\mathcal{S}$ for each dot.
For each boundary pixel $(x,y) \in \mathcal{S}$, we estimate its local background color $\hat{\mathbf{b}}_{x,y}$ by averaging the colors of its neighboring pixels that lie outside the dot region:
\begin{equation*}
\hat{\mathbf{b}}_{x, y} = \frac{1}{|\mathcal{N}_{x, y}|}
\sum_{(u, v) \in \mathcal{N}_{x, y}} \mathbf{I}_{u, v},
\end{equation*}
where $\mathcal{N}_{x,y}$ denotes the eight neighbors of $(x,y)$ that are not part of the dot region (\ie they are neither inner pixels nor boundary pixels).
We then construct an estimated background $\hat{\mathbf{B}}$ by copying $\mathbf{I}$ and replacing the color of every pixel in $\mathcal{S}$ with $\hat{\mathbf{b}}_{x,y}$. 
Note that we do not need to estimate the background for the inner pixels; they are fully covered by the dot color $\mathbf{z}$ during rendering, so their underlying background does not influence the anti-aliasing artifacts and thus does not contribute to location estimation.
The estimated background $\hat{\mathbf{B}}$ then serves
as the input to~\Cref{alg:framework}, and the rest of the recovery process proceeds unchanged.

%% file: tables/alg_desent.tex
\begin{algorithm}[t!]
\caption{\textbf{Automated Location Recovery Framework.} The algorithm
applies perceptual coordinate descent (detailed
in~\Cref{sec:method}) to iteratively refine initial dot locations by
minimizing the perceptual loss computed on anti-aliased boundary pixels
$\mathcal{S}$.}
\label{alg:framework}
\begin{algorithmic}[1]
\REQUIRE Target dot map $\mathbf{I}$, (estimated) background $\mathbf{B}$, map rendering
function $\mathcal{R}$, pixel area of a single dot $\mu$, dot color $\mathbf{z}$,
dot geometry $\phi$, dot size $\rho$, iterations $T$, step size $\eta$
\STATE \textcolor{gray}{\texttt{\# Phase 1: Initialize dot location estimates}}
\STATE $\mathcal{P} \gets
\texttt{FindConnectedComponents}(\mathbf{I}, \mathbf{z})$
\STATE $\mathcal{C} \gets \emptyset$
\algcomment{initialize location}
\FOR{each component $P \in \mathcal{P}$}
    \STATE $k = \lceil \frac{|P|}{\mu} \rceil$
    \algcomment{estimate \# dots in the component}
    \STATE $\mathcal{C}_P \gets
    \texttt{K-Means}(P, k)$
    \STATE $\mathcal{C} \gets \mathcal{C} \cup \mathcal{C}_P$
\ENDFOR

\vspace{0.5em}
\STATE \textcolor{gray}{\texttt{\# Phase 2: Perceptual coordinate
descent (PCD)}}
\myfullcomment{define search directions (stay, left, right, up, down)}
\STATE $\mathcal{D} \gets \{ (0,0), (\eta, 0), (-\eta, 0), (0, \eta),
(0, -\eta) \}$
\myfullcomment{identify boundary pixels}
\STATE $\{\mathcal{S}_i\}_{i=1}^{|\mathcal{C}|} \gets
\texttt{FindBoundaryPixels}(\mathcal{P}, \mathcal{C})$
\FOR{$T$ times}
    \myfullcomment{render candidate maps for each direction}
    \FOR{each $\mathbf{d} \in \mathcal{D}$}
        \STATE $\mathcal{C}_\mathbf{d} \gets \{ \mathbf{c} + \mathbf{d}
        \mid \mathbf{c} \in \mathcal{C} \}$
        \algcomment{shift all locations by $\mathbf{d}$}
        \STATE $\mathbf{I}_\mathbf{d} \gets
        \mathcal{R}(\mathbf{B}, \mathcal{C}_\mathbf{d}, \mathbf{z},
        \phi, \rho)$
        \algcomment{render candidate map}
    \ENDFOR
    \STATE $\mathcal{C}_\text{new} \gets \emptyset$
    \algcomment{initialize updated locations}
    \myfullcomment{evaluate loss for each dot across candidate maps}
    \FOR{each dot $\mathbf{c}_i \in \mathcal{C}$}
        \STATE $\mathbf{c}^\star \gets \mathbf{c}_i +
        \argmin_{\mathbf{d} \in \mathcal{D}}
        \mathcal{L}(\mathbf{I}, \mathbf{I}_\mathbf{d}, \mathbf{B}, \mathcal{S}_i)$
        \STATE $\mathcal{C}_\text{new} \gets
        \mathcal{C}_\text{new} \cup \{ \mathbf{c}^\star \}$
    \ENDFOR
    \STATE $\mathcal{C} \gets \mathcal{C}_\text{new}$
    \algcomment{update all dot locations}
\ENDFOR
\RETURN $\mathcal{C}$
\end{algorithmic}
\end{algorithm}

%% file: 5_0_eval.tex
\section{Evaluation}
\label{sec:exp}

We conduct a comprehensive evaluation of \mymethod across various attack settings to assess the privacy risks associated with different types of dot maps. 
Specifically, we aim to answer the following research questions:

\begin{itemize}
    \item \textbf{RQ1:} How effective is \mymethod compared to existing methods across different configurations of dot maps?
    \item \textbf{RQ2:} How do the different components of \mymethod impact location recovery performance? How efficient is our approach?
    \item \textbf{RQ3:} How does recovery accuracy vary across dots, and what factors contribute to these variations?
\end{itemize}

\input{5_1_setup}

\input{5_2_results}
\input{5_3_ablation}
\input{5_4_insights}

%% file: 5_1_setup.tex
\subsection{Experimental Setup}
\label{sec:exp_setup}

\begin{figure}[t]
  \centering
  \begin{minipage}[]{0.48\linewidth}
    \includegraphics[width=\linewidth]{fig/compressed-openadress-street.png}
    \caption*{(a) OpenAddresses}
  \end{minipage}
  \begin{minipage}[]{0.48\linewidth}
    \includegraphics[width=\linewidth]{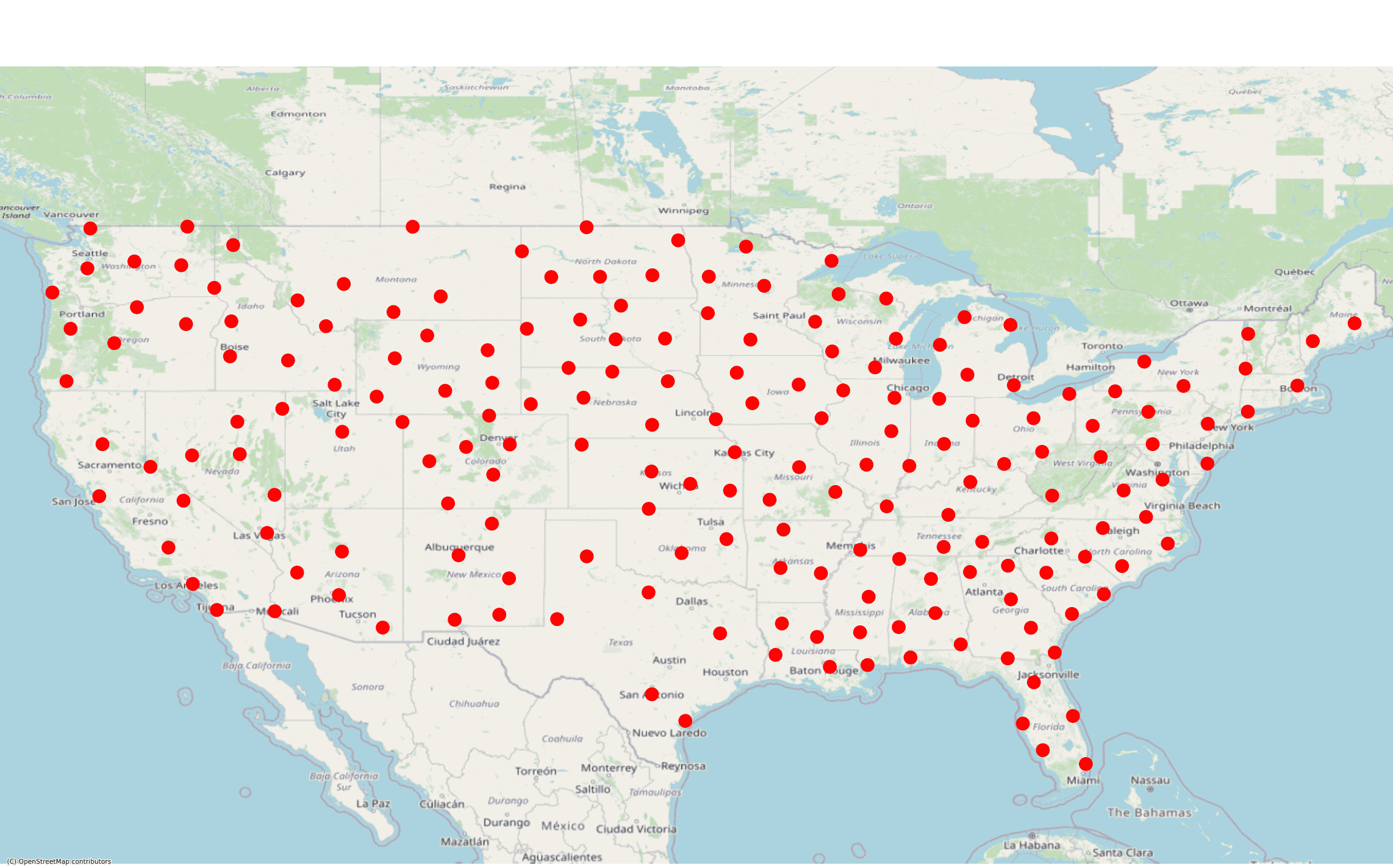}
    \caption*{(b) Synthetic}
  \end{minipage}
  \hfill
  \caption{Dot maps at the small scale (\ie United States) on the OpenAddresses and synthetic datasets.}
  \label{fig:dataset_demo}
\end{figure}

\input{tables/tab_dataset}

\input{tables/tab_map_scale}

\mypara{Evaluation Datasets}
To the best of our knowledge, no public datasets or benchmarks exist for the dot map location recovery task.
Moreover, using dot maps from existing publications would raise privacy concerns and lack ground truth.
To address these issues, we construct two datasets of geographic coordinates for evaluation:
\begin{itemize}
    \item \textit{OpenAddresses.}
    We randomly sample coordinates from OpenAddresses~\cite{openaddresses}, a free and open global collection of geocoded street addresses.
    These coordinates reflect real-world spatial patterns, such as variations in population and urban-rural densities, providing a representative evaluation of privacy risks in practice.
    \item \textit{Synthetic.}
    We uniformly sample coordinates within the geographic range of each map configuration.
    To construct overlapping dots, we randomly perturb the sampled coordinates within a small radius to form clusters of 2, 3, 4, and 5 overlapping dots.
\end{itemize}

Each dataset consists of geographic coordinates (\ie latitude, longitude) that serve as ground-truth locations for the dots.
The OpenAddresses dataset provides coordinates with seven decimal places of latitude and longitude, whereas the synthetic dataset uses six decimal places. 
One unit in the last decimal place corresponds to roughly 1\,cm at seven
decimals and 11\,cm at six decimals.
This level of precision is common in GPS collection systems, and standard geocoding services (\eg Google Maps~\cite{googlemap}) provide six-decimal precision by default. 
Moreover, using this precision ensures that errors in location recovery are attributed to the attack algorithm rather than the coarseness of the input data.
Dataset statistics are summarized in~\Cref{tab:dataset}, and example dot maps generated from these datasets are shown in~\Cref{fig:dataset_demo}.
We further analyze the impact of coordinate precision on attack performance in~\Cref{sec:exp_ablation}.

\mypara{Map Configurations}
To ensure our evaluation covers a realistic range of visualization practices, we systematically vary \textit{seven} map configuration dimensions\fullorcamera{, summarized in~\Cref{tab:map_config}}{, with a summary provided in the full version}.
Specifically, we generate target maps across three geographic scales: small-scale (\ie the United States), medium-scale (\ie Ohio), and large-scale (\ie Austin, Texas).
For each scale, we evaluate three background types: a blank white canvas, a standard street map from OpenStreetMap~\cite{openstreetmap}, and a satellite map provided by Esri~\cite{satellite}.
Maps are exported at three resolutions (96, 192, and 384 DPI), in three image formats (PNG, JPEG, and TIFF), and with three dot geometries (circle, pentagon, and triangle) at varying sizes (1, 2, and 3\,mm).
We use three widely adopted map visualization platforms to generate target dot maps: GeoPandas~\cite{geopandas}, QGIS~\cite{qgis}, and R~\cite{r} (with the maps package~\cite{r_map_pkg}) using their default map composition settings.
A demonstration of the generated small-scale maps across different backgrounds is shown in~\Cref{fig:map_background}.

\mypara{Baselines and Attack Variants}
In addition to the PixelMatch baseline introduced in~\Cref{alg:baseline}, we compare against the following location recovery algorithms:
\begin{itemize}       
    \item \textit{PixelAvg.} This method computes the mean location of both the inner pixels (\ie pixels matching the dot color) and the boundary pixels (\ie pixels adjacent to the inner pixels) to estimate each dot's centroid.
    \item \textit{Raster2Vec.} 
    Image vectorization recovers vector representations from raster images and has been widely studied in computer graphics~\cite{kopf2011depixelizing, pradhan2022vectorgraphics_survey}. 
    As a representative approach, we use QGIS's built-in raster-to-vector conversion tool, which has been applied in prior work~\cite{brownstein2006unsupervised} for location recovery.
    The dot's location is determined as the centroid of the resulting vector geometry.
\end{itemize}

We evaluate both variants of our method: \mymethodBK, which uses perceptual coordinate descent with access to the map background, and \mymethodBU, which estimates the background from the target map when the map background is unavailable.
Existing approaches focus on isolated dots and are designed specifically for that setting. Accordingly, we restrict our comparison with these baselines to the recovery of locations from isolated dots.

\mypara{Evaluation Metrics}
We assess the performance of location recovery algorithms by computing the median recovery error across all dots. 
We report this error using two metrics:
\begin{itemize}
    \item \textit{Absolute Geographical Error.} For each attack, we estimate the dot’s centroid in pixel coordinates and convert it to geographic coordinates (latitude and longitude). 
    We then compute the difference between the estimated and ground-truth locations, reporting latitude error, longitude error, and the geodesic recovery error (Euclidean $L2$ distance) in meters.
    \item \textit{Relative Pixel Error.} We calculate the geographical error (in meters) and normalize it by the real-world distance that a single pixel represents at that map's scale. This provides a relative error at the pixel level.
\end{itemize}

\mypara{Hyperparameter Settings}
We use consistent hyperparameter settings across all experiments for \mymethod to demonstrate its robustness.
Specifically, we set the number of iterations to $T = 30$ and the step size to $\eta = 0.05$ pixels, with learning rate decay~\cite{krogh1991simple} set to $0.75$.
To estimate the per-dot pixel area $\mu$, we randomly select five isolated dots, identify their connected components, and compute the average number of pixels within these components.
For rendering candidate maps, we set the map rendering function $\mathcal{R}$ to the same platform used to generate the target map.
In~\Cref{sec:exp_ablation} we show that performance remains robust when the rendering platform used for the attack differs from the one used to generate the target map.

\mypara{Attack Setup}
We use the Pillow library~\cite{pillow} to load target map images, process pixel data, and run our recovery algorithms.
All evaluated map visualization platforms provide command-line interfaces, enabling automated map generation for our optimization procedure.
We also leverage the coordinate-transformation functions provided by these platforms to convert between pixel and geographic coordinates.
The recovery pipeline is fully automated and requires no human intervention or visual inspection.

%% file: tables/tab_dataset.tex
\begin{table}[t]
    \centering
    \caption{Overview of the evaluation coordinate datasets.}
    \label{tab:dataset}
    \resizebox{0.95\linewidth}{!}{
    \begin{tabular}{lccccccc}
        \toprule
        \multirow{2}{*}{\textbf{Dataset}} & \multicolumn{5}{c}{\textbf{\# Dot Clusters by Overlap}} & \multirow{2}{*}{\textbf{Precision}} & \multirow{2}{*}{\textbf{Type}} \\
        \cmidrule(lr){2-6}
         & 1 (isolated) & 2 & 3 & 4 & 5 & \\
        \midrule
        OpenAddresses & 90 & 24 & 21 & 12 & 5 & 7 decimals & Real-world\\
        Synthetic     & 188 & 25 & 25 & 25 & 25 & 6 decimals & Synthetic\\
        \bottomrule
    \end{tabular}}
\end{table}

%% file: tables/tab_map_scale.tex
\begin{table*}[t]
\centering
\caption{Location recovery error across different map scales. We denote our method variants as \mymethodBK (\ie Background Known) and \mymethodBU (\ie Background Unknown). The best result is in bold.}
\label{tab:map_scale}
\small
\setlength{\tabcolsep}{3pt}
\resizebox{0.94\textwidth}{!}{
\begin{tabular}{cc | cccc | cccc}
\toprule
\multirow{2}{*}{\textbf{Map Scale}} & \multirow{2}{*}{\textbf{Method}}
& \multicolumn{4}{c}{\textbf{OpenAddresses}}
& \multicolumn{4}{c}{\textbf{Synthetic}} \\
\cmidrule(lr){3-6} \cmidrule(lr){7-10}
&
& Lat. Error & Lon. Error & Dist. Error (m) & Rel. Px. Error
& Lat. Error & Lon. Error & Dist. Error (m) & Rel. Px. Error \\
\midrule

\multirow{5}{*}{\begin{tabular}[c]{@{}c@{}}Small-scale \\ (1:10M)\end{tabular}}
& PixelMatch      & 0.002048 & 0.002072 & 287.08 \stdv{105.15} & 0.1063 \stdv{.0389} & 0.001828 & 0.001849 & 256.18 \stdv{97.74} & 0.0949 \stdv{.0362} \\
& PixelAvg        & 0.001685 & 0.001704 & 236.11 \stdv{82.69} & 0.0874 \stdv{.0306} & 0.001617 & 0.001638 & 226.44 \stdv{78.30} & 0.0839 \stdv{.0290} \\
& Raster2Vec      & 0.002048 & 0.002072 & 287.08 \stdv{105.15} & 0.1063 \stdv{.0389} & 0.001828 & 0.001849 & 256.18 \stdv{97.74} & 0.0949 \stdv{.0362} \\
\cmidrule(lr){2-10}
& \mymethodBU     & 0.000020 & 0.000020 & 1.81 \stdv{.66} & 0.0007 \stdv{.0003} & 0.000012 & 0.000012 & 1.69 \stdv{.51} & 0.0006 \stdv{.0002} \\
& \mymethodBK     & \textbf{0.000012} & \textbf{0.000012} & \textbf{0.95} \stdv{.64} & \textbf{0.0003} \stdv{.0002} & \textbf{0.000005} & \textbf{0.000006} & \textbf{0.88} \stdv{.52} & \textbf{0.0002} \stdv{.0001} \\
\midrule

\multirow{5}{*}{\begin{tabular}[c]{@{}c@{}}Medium-scale \\ (1:1M)\end{tabular}}
& PixelMatch      & 0.000219 & 0.000223 & 30.65 \stdv{12.26} & 0.1135 \stdv{.0454} & 0.000274 & 0.000279 & 38.28 \stdv{10.12} & 0.1418 \stdv{.0375} \\
& PixelAvg        & 0.000181 & 0.000185 & 25.41 \stdv{9.78} & 0.0941 \stdv{.0362} & 0.000243 & 0.000247 & 33.96 \stdv{8.18} & 0.1258 \stdv{.0303} \\
& Raster2Vec      & 0.000219 & 0.000223 & 30.65 \stdv{12.26} & 0.1135 \stdv{.0454} & 0.000274 & 0.000279 & 38.28 \stdv{10.12} & 0.1418 \stdv{.0375} \\
\cmidrule(lr){2-10}
& \mymethodBU     & 0.000002 & 0.000002 & 0.25 \stdv{.06} & 0.0009 \stdv{.0002} & 0.000002 & 0.000002 & 0.21 \stdv{.06} & 0.0008 \stdv{.0002} \\
& \mymethodBK     & \textbf{0.000001} & \textbf{0.000001} & \textbf{0.12} \stdv{.06} & \textbf{0.0004} \stdv{.0002} & \textbf{0.000001} & \textbf{0.000001} & \textbf{0.10} \stdv{.05} & \textbf{0.0004} \stdv{.0002} \\
\midrule

\multirow{5}{*}{\begin{tabular}[c]{@{}c@{}}Large-scale \\ (1:100K)\end{tabular}}
& PixelMatch      & 0.000021 & 0.000022 & 2.99 \stdv{1.24} & 0.1107 \stdv{.0460} & 0.000023 & 0.000023 & 3.18 \stdv{1.23} & 0.1178 \stdv{.0454} \\
& PixelAvg        & 0.000019 & 0.000019 & 2.65 \stdv{1.12} & 0.0981 \stdv{.0415} & 0.000017 & 0.000018 & 2.44 \stdv{.97} & 0.0904 \stdv{.0358} \\
& Raster2Vec      & 0.000021 & 0.000022 & 2.99 \stdv{1.24} & 0.1107 \stdv{.0460} & 0.000023 & 0.000023 & 3.18 \stdv{1.23} & 0.1178 \stdv{.0454} \\
\cmidrule(lr){2-10}
& \mymethodBU     & \textbf{0.000001} & \textbf{0.000001} & \textbf{0.05} \stdv{.0046} & \textbf{0.0019} \stdv{.0002} & 0.000001 & 0.000001 & 0.05 \stdv{.0160} & 0.0019 \stdv{.0006} \\
& \mymethodBK     & \textbf{0.000001} & \textbf{0.000001} & \textbf{0.05} \stdv{.0052} & \textbf{0.0019} \stdv{.0002} & \textbf{0.000001} & \textbf{0.000001} & \textbf{0.03} \stdv{.0180} & \textbf{0.0011} \stdv{.0007} \\
\bottomrule
\end{tabular}}
\end{table*}

%% file: 5_2_results.tex
\subsection{Evaluation of \mymethod (RQ1)}
\label{sec:exp_main}

\mypara{Performance Across Different Map Scales}
We use a street map background and GeoPandas to generate target maps at three scales (\ie small, medium, and large) to evaluate location recovery performance.
As shown in~\Cref{tab:map_scale}, all baseline methods (\ie PixelMatch, PixelAvg, and Raster2Vec) produce similar recovery errors on the order of hundreds of meters at the small scale, indicating that they are ineffective at recovering precise locations from dot maps covering broad geographic regions.
In contrast, \mymethodBK and \mymethodBU, which exploit anti-aliasing artifacts, achieve errors of approximately 1~meter and 2~meters on the OpenAddresses dataset, respectively, representing more than $200\times$ and $100\times$ improvements over the strongest baseline.
This demonstrates that precise locations can be recovered even from dot maps covering large geographic areas.
We find that the recovery errors are higher on the OpenAddresses dataset, which we attribute to the complexity of real-world spatial distributions, where higher local dot densities make location estimation more challenging. 
This trend is consistent across scales and datasets, with our methods exhibiting similar relative pixel errors.

\input{tables/tab_background}

\mypara{Performance Across Different Map Backgrounds}
We vary the map background (\ie white canvas, street map, and satellite imagery) to examine its impact on recovery accuracy.
As shown in~\Cref{tab:background_dataset}, baseline methods yield consistently high recovery errors across all backgrounds, as they rely solely on dot color, failing to account for anti-aliasing effects arising from the blending of dots with the underlying background.
In contrast, our proposed methods maintain high location recovery performance across all settings.
Both \mymethodBK and \mymethodBU exhibit a performance drop on highly complex satellite imagery.
This degradation is more pronounced for \mymethodBU, as complex backgrounds reduce the accuracy of local background estimation.
Nevertheless, \mymethodBU still significantly outperforms all baselines, even under these challenging conditions.

\input{tables/tab_resolution}

\mypara{Performance Across Map Resolution}
We further evaluate the impact of image resolution on recovery performance by rendering maps at three different resolutions: 384, 192, and 96 DPI.
The results are shown in~\Cref{tab:resolution}.
As expected, the performance of all methods degrades at lower resolutions due to the reduced number of pixels available to estimate dot centroids accurately.
Despite this, \mymethod maintains strong performance even under low-resolution settings.
At 96 DPI, \mymethodBK still achieves a recovery error of less than 5 meters, which remains sufficient for precise location recovery and continues to significantly outperform the baselines under the same conditions.

\input{tables/tab_gsd}

\mypara{Joint Impact of Map Scale and Resolution}
To study the impact of map scale and resolution jointly, we adopt a single metric, the ground sample distance (GSD), defined as the real-world distance (in meters) represented by one pixel.
We vary the GSD from 27 to 5400\,m/pixel by adjusting the map scale and resolution, and examine the recovery performance on both datasets, as shown in~\Cref{tab:gsd}.
As expected, the distance error of \mymethod grows approximately in proportion to the GSD, since each pixel covers a larger geographic area.
In contrast, the relative pixel error of both methods remains at the millipixel level across all GSDs, indicating that \mymethod recovers dot centroids with a stable sub-pixel precision.

\input{tables/tab_dot_property}

\input{tables/tab_map_tool}

\mypara{Performance Across Dot Properties}
We vary dot geometry and size to examine their impact on recovery accuracy.
As shown in~\Cref{tab:dot_property}, the performance of all methods degrades as dot shapes become more complex (\eg from circles to pentagons).
We observe two distinct trends with respect to dot size. 
First, as the dot size increases, the accuracy of our methods improves, while the performance of the baselines typically deteriorates. 
This is because larger dots introduce more anti-aliased boundary pixels, providing richer sub-pixel information that our optimization can exploit, whereas baseline methods fail to benefit from this additional signal. 
Second, \mymethod remains effective even for very small dots (\ie 1~millimeter in size, corresponding to only about 4~pixels), significantly outperforming the baselines in this challenging regime.

\mypara{Performance Across Map Visualization Platforms}
We evaluate our attack on dot maps generated by three widely used visualization platforms: GeoPandas, QGIS, and R. 
As shown in~\Cref{tab:map_platform}, our methods achieve strong recovery performance, with accuracy within 10 meters.
This indicates that, despite potential differences in underlying (and often unknown) rendering processes, our attack remains robust and achieves high-precision location recovery regardless of the visualization software used.

\mypara{Performance Across Image Formats}
Dot maps are exported in different image formats for dissemination. 
We evaluate our attack on maps saved in three common formats: PNG, TIFF, and JPEG. 
As shown in~\Cref{tab:map_format}, recovery performance is identical for PNG and TIFF, which is expected since both preserve RGB values without compression artifacts.
In contrast, all methods exhibit degraded performance on JPEG images, likely due to compression losses that distort pixel-level information. 
Despite this, our methods still outperform all baselines by nearly $50\times$ under JPEG compression.
Overall, these results demonstrate that our methods generalize well across common raster formats and maintain high-precision location recovery, regardless of the map format.

\input{tables/tab_map_format}
\input{tables/tab_dot_overlap}


\mypara{Performance on Overlapping Dots}
As shown in~\Cref{tab:dot_overlap}, median recovery error generally increases with overlap, since merged regions blur boundaries and make it harder to leverage anti-aliasing artifacts for accurate centroid estimation.
\mymethodBK degrades only slightly in terms of median error, maintaining a median recovery error below 6 meters even with five overlapping dots across both datasets.
In contrast, \mymethodBU shows a larger degradation, with median recovery error increasing from approximately 2 meters for isolated dots to as high as 27 meters under overlap.
Nevertheless, even with five overlapping dots, the median recovery errors of our methods remain substantially lower than those of the baselines on the much simpler task of recovering isolated dots.

\input{tables/tab_region}

\mypara{Performance Across Geographic Regions}
We further evaluate our attack on maps of Mexico to test whether it generalizes beyond the U.S.
Specifically, we sample coordinates from OpenAddresses and generate synthetic data using the same procedure as for the U.S. datasets, and render the maps with all other configurations set to their defaults.
As shown in~\Cref{tab:region}, \mymethodBK recovers locations to within 0.60\,m and 0.73\,m on OpenAddresses and Synthetic, and \mymethodBU is likewise sub-meter.
In contrast, all baselines produce errors on the order of hundreds of meters.
These results show that our attack generalizes across geographic regions.

%% file: tables/tab_background.tex
\begin{table}[t]
\centering
\caption{Location recovery error across map backgrounds.}
\label{tab:background_dataset}
\small
\setlength{\tabcolsep}{3pt}
\resizebox{\columnwidth}{!}{
\begin{tabular}{cc | cc | cc}
\toprule
\multirow{2}{*}{\textbf{Background}} & \multirow{2}{*}{\textbf{Method}}
& \multicolumn{2}{c}{\textbf{OpenAddresses}}
& \multicolumn{2}{c}{\textbf{Synthetic}} \\
\cmidrule(lr){3-4} \cmidrule(lr){5-6}
&
& Dist. Error (m) & Rel. Px. Error
& Dist. Error (m) & Rel. Px. Error \\
\midrule

\multirow{5}{*}{White canvas}
 & PixelMatch          & 287.08 \stdv{105.15} & 0.1063 \stdv{.0389} & 256.18 \stdv{97.74} & 0.0949 \stdv{.0362} \\
 & PixelAvg            & 236.11 \stdv{82.69} & 0.0874 \stdv{.0306} & 226.44 \stdv{78.30} & 0.0839 \stdv{.0290} \\
 & Raster2Vec          & 287.08 \stdv{105.15} & 0.1063 \stdv{.0389} & 256.18 \stdv{97.74} & 0.0949 \stdv{.0362} \\
 \cmidrule(lr){2-6}
 & \mymethodBU         & 1.67 \stdv{.61} & 0.0006 \stdv{.0002} & 1.61 \stdv{.50} & 0.0006 \stdv{.0002} \\
 & \mymethodBK         & \textbf{0.91} \stdv{.62} & \textbf{0.0003} \stdv{.0002} & \textbf{0.81} \stdv{.50} & \textbf{0.0002} \stdv{.0001} \\
\midrule

\multirow{5}{*}{Street}
 & PixelMatch          & 287.08 \stdv{105.15} & 0.1063 \stdv{.0389} & 256.18 \stdv{97.74} & 0.0949 \stdv{.0362} \\
 & PixelAvg            & 236.11 \stdv{82.69} & 0.0874 \stdv{.0306} & 226.44 \stdv{78.30} & 0.0839 \stdv{.0290} \\
 & Raster2Vec          & 287.08 \stdv{105.15} & 0.1063 \stdv{.0389} & 256.18 \stdv{97.74} & 0.0949 \stdv{.0362} \\
 \cmidrule(lr){2-6}
 & \mymethodBU         & 1.81 \stdv{.66} & 0.0007 \stdv{.0003} & 1.69 \stdv{.51} & 0.0006 \stdv{.0002} \\
 & \mymethodBK         & \textbf{0.95} \stdv{.64} & \textbf{0.0003} \stdv{.0002} & \textbf{0.88} \stdv{.52} & \textbf{0.0002} \stdv{.0001} \\
\midrule

\multirow{5}{*}{Satellite}
 & PixelMatch          & 287.08 \stdv{105.15} & 0.1063 \stdv{.0389} & 256.18 \stdv{97.74} & 0.0949 \stdv{.0362} \\
 & PixelAvg            & 236.11 \stdv{82.69} & 0.0874 \stdv{.0306} & 226.44 \stdv{78.30} & 0.0839 \stdv{.0290} \\
 & Raster2Vec          & 287.08 \stdv{105.15} & 0.1063 \stdv{.0389} & 256.18 \stdv{97.74} & 0.0949 \stdv{.0362} \\
 \cmidrule(lr){2-6}
 & \mymethodBU         & 10.50 \stdv{3.64} & 0.0039 \stdv{.0013} & 9.41 \stdv{3.25} & 0.0035 \stdv{.0012} \\
 & \mymethodBK         & \textbf{2.13} \stdv{.67} & \textbf{0.0008} \stdv{.0002} & \textbf{2.73} \stdv{.52} & \textbf{0.0010} \stdv{.0002} \\

\bottomrule
\end{tabular}}
\end{table}

%% file: tables/tab_resolution.tex
\begin{table}[t]
\centering
\caption{Location recovery error across map resolutions.}
\label{tab:resolution}
\small
\setlength{\tabcolsep}{3pt}
\resizebox{\columnwidth}{!}{
\begin{tabular}{cc | cc | cc}
\toprule
\multirow{2}{*}{\textbf{Resolution}} & \multirow{2}{*}{\textbf{Method}}
& \multicolumn{2}{c}{\textbf{OpenAddresses}}
& \multicolumn{2}{c}{\textbf{Synthetic}} \\
\cmidrule(lr){3-4} \cmidrule(lr){5-6}
&
& Dist. Error (m) & Rel. Px. Error
& Dist. Error (m) & Rel. Px. Error \\
\midrule

\multirow{5}{*}{\begin{tabular}[c]{@{}c@{}}384 DPI \\ (4568 $\times$ 2848)\end{tabular}}
 & PixelMatch          & 65.12 \stdv{48.68} & 0.0482 \stdv{.0360} & 64.23 \stdv{48.42} & 0.0476 \stdv{.0359} \\
 & PixelAvg            & 53.11 \stdv{42.45} & 0.0393 \stdv{.0314} & 52.56 \stdv{39.32} & 0.0389 \stdv{.0291} \\
 & Raster2Vec          & 65.12 \stdv{48.68} & 0.0482 \stdv{.0360} & 64.23 \stdv{48.42} & 0.0476 \stdv{.0359} \\
 \cmidrule(lr){2-6}
 & \mymethodBU         & 0.98 \stdv{.52} & 0.0007 \stdv{.0004} & 0.66 \stdv{.28} & 0.0005 \stdv{.0002} \\
 & \mymethodBK         & \textbf{0.46} \stdv{.33} & \textbf{0.0003} \stdv{.0002} & \textbf{0.36} \stdv{.31} & \textbf{0.0003} \stdv{.0003} \\
\midrule

\multirow{5}{*}{\begin{tabular}[c]{@{}c@{}}192 DPI \\ (2284 $\times$ 1424)\end{tabular}}
 & PixelMatch          & 287.08 \stdv{105.15} & 0.1063 \stdv{.0389} & 256.18 \stdv{97.74} & 0.0949 \stdv{.0362} \\
 & PixelAvg            & 236.11 \stdv{82.69} & 0.0874 \stdv{.0306} & 226.44 \stdv{78.30} & 0.0839 \stdv{.0290} \\
 & Raster2Vec          & 287.08 \stdv{105.15} & 0.1063 \stdv{.0389} & 256.18 \stdv{97.74} & 0.0949 \stdv{.0362} \\
 \cmidrule(lr){2-6}
 & \mymethodBU         & 1.81 \stdv{.66} & 0.0007 \stdv{.0003} & 1.69 \stdv{.51} & 0.0006 \stdv{.0002} \\
 & \mymethodBK         & \textbf{0.95} \stdv{.64} & \textbf{0.0003} \stdv{.0002} & \textbf{0.88} \stdv{.52} & \textbf{0.0002} \stdv{.0001} \\
\midrule

\multirow{5}{*}{\begin{tabular}[c]{@{}c@{}}96 DPI \\ (1142 $\times$ 712)\end{tabular}}
& PixelMatch   & 370.67 \stdv{224.32} & 0.0686 \stdv{.0415} & 335.56 \stdv{186.80} & 0.0621 \stdv{.0346} \\
& PixelAvg     & 405.33 \stdv{192.50} & 0.0751 \stdv{.0357} & 375.08 \stdv{144.41} & 0.0695 \stdv{.0268} \\
& Raster2Vec   & 370.67 \stdv{224.32} & 0.0686 \stdv{.0415} & 335.56 \stdv{186.80} & 0.0621 \stdv{.0346} \\
 \cmidrule(lr){2-6}
 & \mymethodBU         & 7.36 \stdv{3.07} & 0.0014 \stdv{.0006} & 6.84 \stdv{2.04} & 0.0013 \stdv{.0004} \\
 & \mymethodBK         & \textbf{4.28} \stdv{1.64} & \textbf{0.0008} \stdv{.0003} & \textbf{3.84} \stdv{1.06} & \textbf{0.0007} \stdv{.0002} \\
\bottomrule
\end{tabular}
}
\end{table}

%% file: tables/tab_gsd.tex
\begin{table}[t]
\centering
\caption{Location recovery error across different ground sample distances (GSD).}
\label{tab:gsd}
\small
\setlength{\tabcolsep}{3pt}
\resizebox{\columnwidth}{!}{
\begin{tabular}{cc | cc | cc}
\toprule
\multirow{2}{*}{\begin{tabular}[c]{@{}c@{}}\textbf{GSD}\\ \textbf{(m/pixel)}\end{tabular}}
& \multirow{2}{*}{\textbf{Method}}
& \multicolumn{2}{c}{\textbf{OpenAddresses}}
& \multicolumn{2}{c}{\textbf{Synthetic}} \\
\cmidrule(lr){3-4}
\cmidrule(lr){5-6}
&
& Dist. Error (m) & Rel. Px. Error
& Dist. Error (m) & Rel. Px. Error \\
\midrule
\multirow{2}{*}{27}
& \mymethodBU & 0.05 \stdv{.0046} & 0.0019 \stdv{.0002} & 0.05 \stdv{.0160} & 0.0019 \stdv{.0006} \\
& \mymethodBK & 0.05 \stdv{.0052} & 0.0019 \stdv{.0002} & 0.03 \stdv{.0180} & 0.0011 \stdv{.0007} \\
\midrule
\multirow{2}{*}{270}
& \mymethodBU & 0.25 \stdv{.06} & 0.0009 \stdv{.0002} & 0.21 \stdv{.06} & 0.0008 \stdv{.0002} \\
& \mymethodBK & 0.12 \stdv{.06} & 0.0004 \stdv{.0002} & 0.10 \stdv{.05} & 0.0004 \stdv{.0002} \\
\midrule
\multirow{2}{*}{1350}
& \mymethodBU & 0.98 \stdv{.52} & 0.0007 \stdv{.0004} & 0.66 \stdv{.28} & 0.0005 \stdv{.0002} \\
& \mymethodBK & 0.46 \stdv{.33} & 0.0003 \stdv{.0002} & 0.36 \stdv{.31} & 0.0003 \stdv{.0003} \\
\midrule
\multirow{2}{*}{2700}
& \mymethodBU & 1.81 \stdv{.66} & 0.0007 \stdv{.0003} & 1.69 \stdv{.51} & 0.0006 \stdv{.0002} \\
& \mymethodBK & 0.95 \stdv{.64} & 0.0003 \stdv{.0002} & 0.88 \stdv{.52} & 0.0002 \stdv{.0001} \\
\midrule
\multirow{2}{*}{5400}
& \mymethodBU & 7.36 \stdv{3.07} & 0.0014 \stdv{.0006} & 6.84 \stdv{2.04} & 0.0013 \stdv{.0004} \\
& \mymethodBK & 4.28 \stdv{1.64} & 0.0008 \stdv{.0003} & 3.84 \stdv{1.06} & 0.0007 \stdv{.0002} \\
\bottomrule
\end{tabular}
}
\end{table}

%% file: tables/tab_dot_property.tex
\begin{table}[t]
\centering
\caption{Comparison of the location recovery error (in meters) across different dot geometries and sizes.}
\label{tab:dot_property}
\resizebox{0.99\linewidth}{!}{
\begin{tabular}{cc | ccc | ccc}
\toprule
\multirow{2}{*}{\textbf{Geometry}} & \multirow{2}{*}{\textbf{Method}}
& \multicolumn{3}{c}{\textbf{OpenAddresses}}
& \multicolumn{3}{c}{\textbf{Synthetic}} \\
\cmidrule(lr){3-5} \cmidrule(lr){6-8}
& & 1mm & 2mm & 3mm & 1mm & 2mm & 3mm \\
\midrule

\multirow{5}{*}{Circle}
 & PixelMatch          & 168.05 \stdv{149.42} & 287.08 \stdv{105.15} & 99.56 \stdv{68.75} & 181.08 \stdv{105.24} & 256.18 \stdv{97.74} & 126.84 \stdv{62.93} \\
 & PixelAvg            & 219.19 \stdv{134.29} & 236.11 \stdv{82.69} & 103.85 \stdv{55.04} & 196.25 \stdv{112.78} & 226.44 \stdv{78.30} & 102.16 \stdv{49.52} \\
 & Raster2Vec          & 168.05 \stdv{149.42} & 287.08 \stdv{105.15} & 99.56 \stdv{68.75} & 181.08 \stdv{105.24} & 256.18 \stdv{97.74} & 126.84 \stdv{62.93} \\
 \cmidrule(lr){2-8}
 & \mymethodBU         & 7.68 \stdv{4.98} & 1.81 \stdv{.66} & 1.80 \stdv{1.15}
                       & 5.33 \stdv{3.30} & 1.69 \stdv{.51} & 1.60 \stdv{1.12} \\
 & \mymethodBK         & \textbf{1.73} \stdv{1.11} & \textbf{0.95} \stdv{.64} & \textbf{0.94} \stdv{.64}
                       & \textbf{1.92} \stdv{1.16} & \textbf{0.88} \stdv{.52} & \textbf{0.87} \stdv{.57} \\

\midrule

\multirow{5}{*}{Pentagon}
 & PixelMatch          & 271.80 \stdv{163.90} & 426.26 \stdv{181.30} & 316.41 \stdv{190.00} & 300.77 \stdv{151.09} & 323.57 \stdv{171.63} & 306.25 \stdv{177.14} \\
 & PixelAvg            & 263.62 \stdv{162.24} & 392.14 \stdv{180.89} & 289.59 \stdv{176.50} & 293.30 \stdv{165.56} & 285.61 \stdv{174.03} & 265.63 \stdv{164.79} \\
 & Raster2Vec          & 271.80 \stdv{163.90} & 426.26 \stdv{181.30} & 316.41 \stdv{190.00} & 300.77 \stdv{151.09} & 323.57 \stdv{171.63} & 306.25 \stdv{177.14} \\
 \cmidrule(lr){2-8}
 & \mymethodBU         & 7.90 \stdv{5.63} & 6.61 \stdv{5.43} & 5.55 \stdv{2.88}
                       & 10.49 \stdv{7.33} & 5.17 \stdv{2.71} & 4.47 \stdv{2.93} \\
 & \mymethodBK         & \textbf{4.04} \stdv{2.51} & \textbf{3.97} \stdv{2.96} & \textbf{2.25} \stdv{.80}
                       & \textbf{4.76} \stdv{2.89} & \textbf{3.70} \stdv{2.11} & \textbf{2.24} \stdv{1.57} \\

\midrule

\multirow{5}{*}{Triangle}
 & PixelMatch          & 441.29 \stdv{264.24} & 392.82 \stdv{200.83} & 513.14 \stdv{292.00} & 477.40 \stdv{281.77} & 406.01 \stdv{262.56} & 424.83 \stdv{254.01} \\
 & PixelAvg            & 466.58 \stdv{303.88} & 391.55 \stdv{192.16} & 463.76 \stdv{288.14} & 442.65 \stdv{284.13} & 387.77 \stdv{266.23} & 398.10 \stdv{267.11} \\
 & Raster2Vec          & 441.29 \stdv{264.24} & 392.82 \stdv{200.83} & 513.14 \stdv{292.00} & 477.40 \stdv{281.77} & 406.01 \stdv{262.56} & 424.83 \stdv{254.01} \\
 \cmidrule(lr){2-8}
 & \mymethodBU         & 15.18 \stdv{8.81} & 8.56 \stdv{5.04} & 7.68 \stdv{4.04}
                       & 10.72 \stdv{6.41} & 7.43 \stdv{4.10} & 6.17 \stdv{3.42} \\
 & \mymethodBK         & \textbf{10.22} \stdv{5.30} & \textbf{6.78} \stdv{3.94} & \textbf{4.58} \stdv{2.39}
                       & \textbf{8.21} \stdv{4.78} & \textbf{5.86} \stdv{3.48} & \textbf{4.09} \stdv{2.31} \\

\bottomrule
\end{tabular}}
\end{table}

%% file: tables/tab_map_tool.tex
\begin{table}[t]
\centering
\caption{Location recovery error across map tools.}
\label{tab:map_platform}
\small
\setlength{\tabcolsep}{3pt}
\resizebox{\columnwidth}{!}{
\begin{tabular}{cc | cc | cc}
\toprule
\multirow{2}{*}{\textbf{Platform}} & \multirow{2}{*}{\textbf{Method}}
& \multicolumn{2}{c}{\textbf{OpenAddresses}}
& \multicolumn{2}{c}{\textbf{Synthetic}} \\
\cmidrule(lr){3-4} \cmidrule(lr){5-6}
& & Dist. Error (m) & Rel. Px. Error
  & Dist. Error (m) & Rel. Px. Error \\
\midrule

\multirow{5}{*}{GeoPandas}
 & PixelMatch          & 287.08 \stdv{105.15} & 0.1063 \stdv{.0389} & 256.18 \stdv{97.74} & 0.0949 \stdv{.0362} \\
 & PixelAvg            & 236.11 \stdv{82.69} & 0.0874 \stdv{.0306} & 226.44 \stdv{78.30} & 0.0839 \stdv{.0290} \\
 & Raster2Vec          & 287.08 \stdv{105.15} & 0.1063 \stdv{.0389} & 256.18 \stdv{97.74} & 0.0949 \stdv{.0362} \\
 \cmidrule(lr){2-6}
 & \mymethodBU         & 1.81 \stdv{.66} & 0.0007 \stdv{.0003} & 1.69 \stdv{.51} & 0.0006 \stdv{.0002} \\
 & \mymethodBK         & \textbf{0.95} \stdv{.64} & \textbf{0.0003} \stdv{.0002} & \textbf{0.88} \stdv{.52} & \textbf{0.0002} \stdv{.0001} \\
\midrule

\multirow{5}{*}{QGIS}
& PixelMatch          & 258.13 \stdv{99.80} & 0.0956 \stdv{.0367} & 227.74 \stdv{82.40} & 0.0844 \stdv{.0308} \\
& PixelAvg            & 291.12 \stdv{96.40} & 0.1078 \stdv{.0355} & 273.30 \stdv{98.70} & 0.1012 \stdv{.0368} \\
& Raster2Vec          & 258.13 \stdv{99.80} & 0.0956 \stdv{.0367} & 227.74 \stdv{82.40} & 0.0844 \stdv{.0308} \\
\cmidrule(lr){2-6}
& \mymethodBU         & 9.51 \stdv{3.21} & 0.0035 \stdv{.0012} & 8.01 \stdv{2.70} & 0.0030 \stdv{.0010} \\
& \mymethodBK         & \textbf{9.14} \stdv{5.54} & \textbf{0.0034} \stdv{.0020} & \textbf{2.86} \stdv{1.91} & \textbf{0.0011} \stdv{.0007} \\
\midrule

\multirow{5}{*}{R}
& PixelMatch          & 262.53 \stdv{91.70} & 0.0972 \stdv{.0342} & 261.67 \stdv{105.60} & 0.0969 \stdv{.0390} \\
& PixelAvg            & 291.76 \stdv{109.30} & 0.1080 \stdv{.0406} & 220.71 \stdv{72.50} & 0.0817 \stdv{.0269} \\
& Raster2Vec          & 262.53 \stdv{91.70} & 0.0972 \stdv{.0342} & 261.67 \stdv{105.60} & 0.0969 \stdv{.0390} \\
\cmidrule(lr){2-6}
& \mymethodBU         & 4.01 \stdv{1.58} & 0.0015 \stdv{.0006} & 3.93 \stdv{1.08} & 0.0015 \stdv{.0004} \\
& \mymethodBK         & \textbf{2.58} \stdv{1.61} & \textbf{0.0010} \stdv{.0006} & \textbf{2.50} \stdv{1.57} & \textbf{0.0009} \stdv{.0006} \\
\bottomrule
\end{tabular}
}
\end{table}

%% file: tables/tab_map_format.tex
\begin{table}[t]
\centering
\caption{Location recovery error across map formats.}
\label{tab:map_format}
\resizebox{\linewidth}{!}{
\begin{tabular}{cc | cc | cc}
\toprule
\multirow{2}{*}{\textbf{Map Format}} & \multirow{2}{*}{\textbf{Method}} 
& \multicolumn{2}{c}{\textbf{OpenAddresses}} 
& \multicolumn{2}{c}{\textbf{Synthetic}} \\
\cmidrule(lr){3-4} \cmidrule(lr){5-6}
& & Dist. Error (m) & Rel. Px. Error 
  & Dist. Error (m) & Rel. Px. Error \\
\midrule

\multirow{5}{*}{PNG}
 & PixelMatch          & 287.08 \stdv{105.15} & 0.1063 \stdv{.0389} & 256.18 \stdv{97.74} & 0.0949 \stdv{.0362} \\
 & PixelAvg            & 236.11 \stdv{82.69} & 0.0874 \stdv{.0306} & 226.44 \stdv{78.30} & 0.0839 \stdv{.0290} \\
 & Raster2Vec          & 287.08 \stdv{105.15} & 0.1063 \stdv{.0389} & 256.18 \stdv{97.74} & 0.0949 \stdv{.0362} \\
 \cmidrule(lr){2-6}
 & \mymethodBU         & 1.81 \stdv{.66} & 0.0007 \stdv{.0003} & 1.69 \stdv{.51} & 0.0006 \stdv{.0002} \\
 & \mymethodBK         & \textbf{0.95} \stdv{.64} & \textbf{0.0003} \stdv{.0002} & \textbf{0.88} \stdv{.52} & \textbf{0.0002} \stdv{.0001} \\
\midrule

\multirow{5}{*}{TIFF}
 & PixelMatch          & 287.08 \stdv{105.15} & 0.1063 \stdv{.0389} & 256.18 \stdv{97.74} & 0.0949 \stdv{.0362} \\
 & PixelAvg            & 236.11 \stdv{82.69} & 0.0874 \stdv{.0306} & 226.44 \stdv{78.30} & 0.0839 \stdv{.0290} \\
 & Raster2Vec          & 287.08 \stdv{105.15} & 0.1063 \stdv{.0389} & 256.18 \stdv{97.74} & 0.0949 \stdv{.0362} \\
 \cmidrule(lr){2-6}
 & \mymethodBU         & 1.81 \stdv{.66} & 0.0007 \stdv{.0003} & 1.69 \stdv{.51} & 0.0006 \stdv{.0002} \\
 & \mymethodBK         & \textbf{0.95} \stdv{.64} & \textbf{0.0003} \stdv{.0002} & \textbf{0.88} \stdv{.52} & \textbf{0.0002} \stdv{.0001} \\

\midrule

\multirow{5}{*}{JPEG}
 & PixelMatch          & 320.53 \stdv{88.07} & 0.1187 \stdv{.0326} & 333.03 \stdv{91.79} & 0.1233 \stdv{.0340} \\
 & PixelAvg            & 308.76 \stdv{60.77} & 0.1144 \stdv{.0225} & 294.37 \stdv{66.22} & 0.1090 \stdv{.0245} \\
 & Raster2Vec          & 320.53 \stdv{88.07} & 0.1187 \stdv{.0326} & 333.03 \stdv{91.79} & 0.1233 \stdv{.0340} \\
 \cmidrule(lr){2-6}
 & \mymethodBU         & 7.46 \stdv{26.87} & 0.0028 \stdv{.0072} & 7.12 \stdv{6.68} & 0.0026 \stdv{.0024} \\
 & \mymethodBK         & \textbf{6.27} \stdv{5.09} & \textbf{0.0023} \stdv{.0018} & \textbf{5.84} \stdv{5.17} & \textbf{0.0022} \stdv{.0019} \\
\bottomrule
\end{tabular}}
\end{table}

%% file: tables/tab_dot_overlap.tex
\begin{table}[t]
\centering
\caption{Location recovery error of isolated dots and overlapping dots.}
\label{tab:dot_overlap}
\resizebox{\linewidth}{!}{
\begin{tabular}{cc | cc | cc}
\toprule
\multirow{2}{*}{\textbf{\# Overlaps}} & \multirow{2}{*}{\textbf{Method}} 
& \multicolumn{2}{c}{\textbf{OpenAddresses}} 
& \multicolumn{2}{c}{\textbf{Synthetic}} \\
\cmidrule(lr){3-4} \cmidrule(lr){5-6}
& & Dist. Error (m) & Rel. Px. Error 
  & Dist. Error (m) & Rel. Px. Error \\
\midrule

\multirow{5}{*}{1}
 & PixelMatch          & 287.08 \stdv{105.15} & 0.1063 \stdv{.0389} & 256.18 \stdv{97.74} & 0.0949 \stdv{.0362} \\
 & PixelAvg            & 236.11 \stdv{82.69} & 0.0874 \stdv{.0306} & 226.44 \stdv{78.30} & 0.0839 \stdv{.0290} \\
 & Raster2Vec          & 287.08 \stdv{105.15} & 0.1063 \stdv{.0389} & 256.18 \stdv{97.74} & 0.0949 \stdv{.0362} \\
 \cmidrule(lr){2-6}
 & \mymethodBU         & 1.81 \stdv{.66} & 0.0007 \stdv{.0003} & 1.69 \stdv{.51} & 0.0006 \stdv{.0002} \\
 & \mymethodBK         & \textbf{0.95} \stdv{.64} & \textbf{0.0003} \stdv{.0002} & \textbf{0.88} \stdv{.52} & \textbf{0.0002} \stdv{.0001} \\

\midrule

\multirow{2}{*}{2}
 & \mymethodBU         & 7.34 \stdv{102.29} & 0.0027 \stdv{.0376} & 3.00 \stdv{1.66} & 0.0011 \stdv{.0005} \\
 & \mymethodBK         & \textbf{2.26} \stdv{1.04} & \textbf{0.0008} \stdv{.0004} & \textbf{1.82} \stdv{.70} & \textbf{0.0007} \stdv{.0003} \\

\midrule

\multirow{2}{*}{3}
 & \mymethodBU         & 8.89 \stdv{185.73} & 0.0033 \stdv{.0689} & 4.79 \stdv{50.39} & 0.0018 \stdv{.0189} \\
 & \mymethodBK         & \textbf{4.31} \stdv{32.54} & \textbf{0.0016} \stdv{.0121} & \textbf{2.17} \stdv{.99} & \textbf{0.0008} \stdv{.0004} \\

\midrule

\multirow{2}{*}{4}
 & \mymethodBU         & 27.00 \stdv{1046.78} & 0.0100 \stdv{.3877} & 4.39 \stdv{161.92} & 0.0016 \stdv{.0590} \\
 & \mymethodBK         & \textbf{4.86} \stdv{241.87} & \textbf{0.0018} \stdv{.0896} & \textbf{2.50} \stdv{15.54} & \textbf{0.0009} \stdv{.0056} \\

\midrule

\multirow{2}{*}{5}
 & \mymethodBU         & 16.12 \stdv{11828.85} & 0.0060 \stdv{4.4028} & 5.37 \stdv{641.73} & 0.0020 \stdv{.2390} \\
 & \mymethodBK         & \textbf{5.46} \stdv{10264.44} & \textbf{0.0020} \stdv{3.7599} & \textbf{2.92} \stdv{1158.51} & \textbf{0.0011} \stdv{.4364} \\

\bottomrule
\end{tabular}}
\end{table}

%% file: tables/tab_region.tex
\begin{table}[t]
\centering
\caption{Location recovery performance in Mexico.}
\label{tab:region}
\small
\setlength{\tabcolsep}{3pt}
\resizebox{\columnwidth}{!}{
\begin{tabular}{c | cc | cc}
\toprule
\multirow{2}{*}{\textbf{Method}}
& \multicolumn{2}{c}{\textbf{OpenAddresses}}
& \multicolumn{2}{c}{\textbf{Synthetic}} \\
\cmidrule(lr){2-3} \cmidrule(lr){4-5}
& Dist. Error (m) & Rel. Px. Error
& Dist. Error (m) & Rel. Px. Error \\
\midrule
PixelMatch  & 157.34 \stdv{51.00} & 0.1075 \stdv{.0348} & 162.74 \stdv{55.54} & 0.1112 \stdv{.0379} \\
PixelAvg    & 132.43 \stdv{39.52} & 0.0905 \stdv{.0270} & 141.70 \stdv{46.62} & 0.0968 \stdv{.0319} \\
Raster2Vec  & 157.34 \stdv{51.00} & 0.1075 \stdv{.0348} & 162.74 \stdv{55.54} & 0.1112 \stdv{.0379} \\
\midrule
\mymethodBU & 0.69 \stdv{.33} & 0.0005 \stdv{.0002} & 0.76 \stdv{.34} & 0.0005 \stdv{.0002} \\
\mymethodBK & \textbf{0.60} \stdv{.35} & \textbf{0.0004} \stdv{.0002} & \textbf{0.73} \stdv{.31} & \textbf{0.0005} \stdv{.0002} \\
\bottomrule
\end{tabular}}
\end{table}

%% file: 5_3_ablation.tex
\subsection{Ablation Study (RQ2)}
\label{sec:exp_ablation}

\begin{figure}[t]
    \centering
    \includegraphics[width=0.43\textwidth]{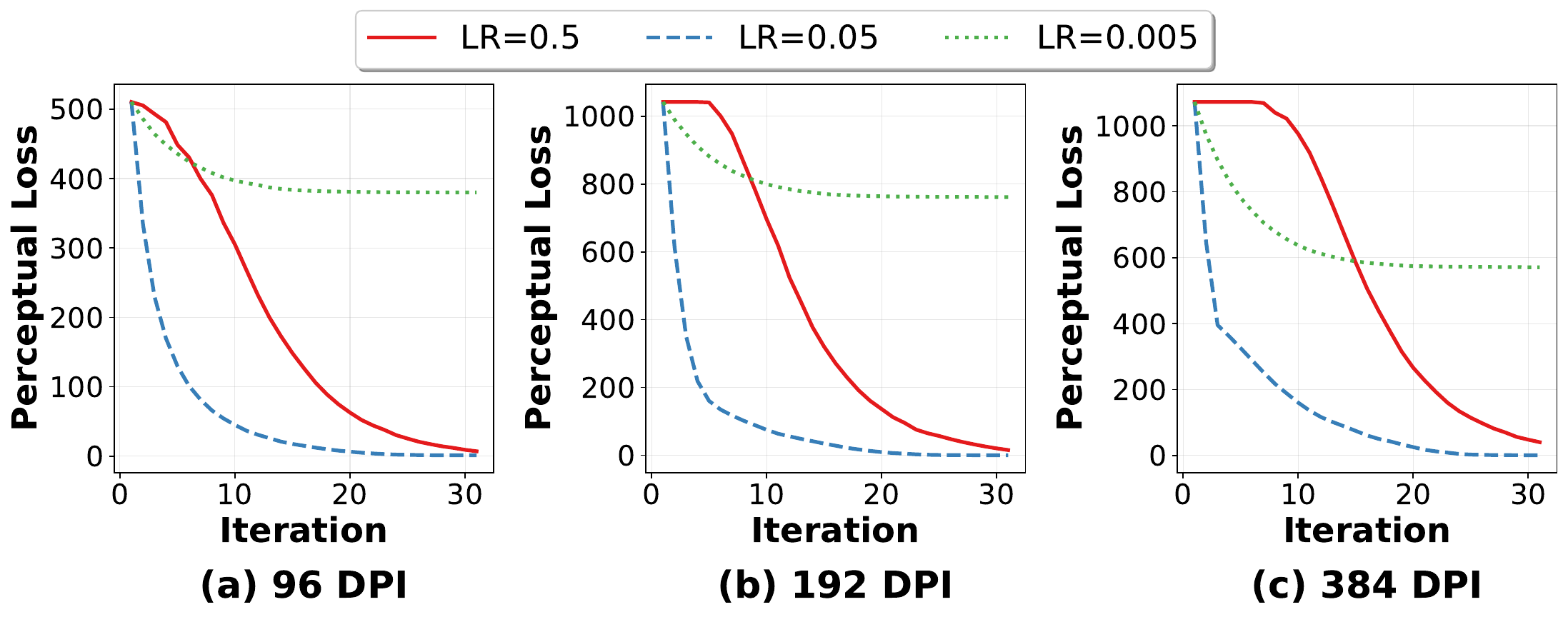}
    \caption{Impact of iterations and step size in \mymethodBK.}
    \label{fig:loss}
\end{figure}

\mypara{Impact of Optimization Hyperparameters}
We analyze the effect of the optimization iterations $T$ and step size $\eta$ on the convergence and recovery performance of \mymethodBK (results for \mymethodBU are omitted due to similar trends).
\Cref{fig:loss} shows the perceptual loss trajectories at different resolutions under varying step sizes.
We find that a large step size (\ie 0.5) fails to converge and yields inaccurate estimates, while a small step size (\ie 0.005) converges too slowly. 
Overall, our default setting (\ie $\eta=0.05$) converges reliably within 30 iterations and achieves the lowest perceptual loss. 
These results are consistent across map resolutions, highlighting the stability of our algorithm.

\mypara{Impact of Background in \mymethodBK}
In \mymethodBK, perceptual loss is computed by rendering a new map using the available background and the currently estimated dot centroids. 
In practice, the adversary may not have access to the exact background used in the target map. 
We therefore evaluate how background variations affect recovery performance.
Specifically, we use a street map from OpenStreetMap as the ground-truth background and consider three scenarios:
(i) \textit{Same}: the background is identical to that of the target map;
(ii) \textit{Similar}: a different street map of the same region from Esri~\cite{esrimap}, captured in a different year;
(iii) \textit{Perturbed}: the background is corrupted with additive Gaussian noise (\ie $\mathcal{N}(0,\sigma^2)$) applied independently to each RGB channel (with $\sigma$ varying from 5 to 25), while preserving overall visual appearance.

The results are shown in~\Cref{tab:background_impact}. 
We find that even small deviations from the true background (\eg perturbations with $\sigma=5$) lead to noticeable drops in recovery accuracy. 
This arises because \mymethodBK depends on subtle anti-aliasing artifacts along dot boundaries; any discrepancy in the background disrupts these cues, resulting in inaccurate loss evaluation and suboptimal optimization.
When the background differs substantially from the ground truth (\eg large perturbations or different map sources), the performance of \mymethodBK falls below that of \mymethodBU, which directly estimates the background from the target map. 
In such cases, we recommend using \mymethodBU when the background is unknown or cannot be accurately reproduced.

\mypara{Impact of Map Visualization Tools Used in \mymethod}
In our previous experiments, we assumed that the adversary has access to the same visualization tool (\ie rendering function $\mathcal{R}$) used to generate the target map. 
In practice, however, this assumption may not hold.
To evaluate this scenario, we fix the target map to one generated using GeoPandas with a street map background and vary the rendering tool used by the adversary during optimization. 
The results in~\Cref{tab:impact_render} show that mismatches between the target and adversary tools have only a minor impact on recovery accuracy, with location recovery errors remaining within 3~meters for both \mymethodBK and \mymethodBU.
This demonstrates that \mymethod is robust even when the adversary does not know, or have access to, the specific tool used to generate the target map.

\input{tables/tab_diff_background}

\input{tables/tab_diff_render}

\input{tables/tab_efficiency}

\input{tables/tab_decimals}

\mypara{Efficiency Evaluation}
We evaluate the efficiency of our recovery algorithms and compare them with the baselines on a laptop equipped with a Snapdragon X Elite CPU.
As shown in~\Cref{tab:running_time}, the baseline methods are highly efficient, requiring less than one minute to estimate dot centroids across different resolutions.
Our methods (\mymethodBK and \mymethodBU) incur a higher computational cost, taking several minutes to complete. 
We find that this overhead is dominated by repeated map rendering during optimization, particularly at higher resolutions.
Nevertheless, the overall runtime remains practical, requiring only a few minutes on a laptop.
These results demonstrate that our attacks are feasible for adversaries without access to specialized computational resources.

\mypara{Impact of Input Coordinate Precision}
To evaluate whether reduced input coordinate precision affects attack performance, we vary the decimal precision of the input coordinates from 4 to 6 digits when generating the dot maps (where 4 decimal places correspond to approximately 11 meters of spatial precision). 
For each precision level, we generate target maps and measure recovery error with respect to the original high-precision coordinates.
As shown in~\Cref{tab:eval_decimals}, reducing precision to 4 or 5 decimal places has only a modest impact on recovery accuracy, with errors remaining below 6~meters across both datasets. 
This still significantly outperforms all baselines, which incur errors on the order of hundreds of meters.
These results demonstrate that \mymethod remains effective even when the input coordinates are provided at coarse precision.

\input{tables/tab_boundary}

%

\mypara{Impact of Boundary Pixel Selection}
By default, \mymethod computes the perceptual loss over the boundary pixels $\mathcal{S}$, \ie the four-neighbors of the inner pixels.
To evaluate whether \mymethod can benefit from pixels farther from the dot, we extend $\mathcal{S}$ by including the four-neighbors of the boundary pixels themselves (excluding inner pixels), and compare the resulting recovery performance.
As shown in~\Cref{tab:boundary}, we observe no noticeable performance difference, suggesting that the default four-neighbor boundary set is sufficient for accurate location recovery with \mymethod.

%% file: tables/tab_diff_background.tex
\begin{table}[t]
\centering
\caption{Impact of different map backgrounds on the location recovery error of \mymethodBK.}
\label{tab:background_impact}
\resizebox{0.95\linewidth}{!}{
\begin{tabular}{lcccc}
\toprule
\multirow{2}{*}{\textbf{Background}} & \multicolumn{2}{c}{\textbf{OpenAddresses}} & \multicolumn{2}{c}{\textbf{Synthetic}} \\
\cmidrule(lr){2-3} \cmidrule(lr){4-5}
& \textbf{Dist. Error (m)} & \textbf{Rel. Px. Error} & \textbf{Dist. Error (m)} & \textbf{Rel. Px. Error} \\
\midrule
Same  & 0.95 \stdv{.64} & 0.0003 \stdv{.0002} & 0.88 \stdv{.52} & 0.0002 \stdv{.0001} \\
Similar & 8.92 \stdv{5.75} & 0.0033 \stdv{.0021} & 8.62 \stdv{5.39} & 0.0032 \stdv{.0020} \\
Perturbed ($\sigma=5$)  & 11.37 \stdv{7.39} & 0.0042 \stdv{.0027} & 10.93 \stdv{6.89} & 0.0040 \stdv{.0026} \\
Perturbed ($\sigma=10$) & 19.48 \stdv{12.37} & 0.0072 \stdv{.0046} & 18.64 \stdv{12.21} & 0.0069 \stdv{.0045} \\
Perturbed ($\sigma=15$) & 29.86 \stdv{19.71} & 0.0111 \stdv{.0072} & 27.04 \stdv{17.31} & 0.0100 \stdv{.0065} \\
Perturbed ($\sigma=20$) & 28.87 \stdv{18.62} & 0.0107 \stdv{.0070} & 31.35 \stdv{19.59} & 0.0116 \stdv{.0071} \\
Perturbed ($\sigma=25$) & 35.04 \stdv{22.78} & 0.0130 \stdv{.0083} & 35.29 \stdv{22.59} & 0.0131 \stdv{.0083} \\
\bottomrule
\end{tabular}}
\end{table}

%% file: tables/tab_diff_render.tex
\begin{table}[t]
\centering
\caption{Impact of rendering platform for \mymethod across datasets. The target map is generated using GeoPandas.}
\label{tab:impact_render}
\resizebox{\linewidth}{!}{
\begin{tabular}{cc | cc | cc}
\toprule
\multirow{2}{*}{\textbf{Platform}} & \multirow{2}{*}{\textbf{Method}}
& \multicolumn{2}{c}{\textbf{OpenAddresses}}
& \multicolumn{2}{c}{\textbf{Synthetic}} \\
\cmidrule(lr){3-4} \cmidrule(lr){5-6}
& & Dist. Error (m) & Rel. Px. Error
  & Dist. Error (m) & Rel. Px. Error \\
\midrule

\multirow{2}{*}{GeoPandas}
 & \mymethodBU       & 1.81 \stdv{.66} & 0.0007 \stdv{.0003} & 1.69 \stdv{.51} & 0.0006 \stdv{.0002} \\
 & \mymethodBK       & 0.95 \stdv{.64} & 0.0003 \stdv{.0002} & 0.88 \stdv{.52} & 0.0002 \stdv{.0001}\\
\midrule

\multirow{2}{*}{QGIS}
 & \mymethodBU       & 2.67 \stdv{1.21} & 0.0010 \stdv{.0006} & 2.36 \stdv{1.70} & 0.0009 \stdv{.0007} \\
 & \mymethodBK       & 2.35 \stdv{1.54} & 0.0009 \stdv{.0006} & 2.41 \stdv{1.91} & 0.0009 \stdv{.0007} \\
\midrule

\multirow{2}{*}{R}
 & \mymethodBU       & 2.51 \stdv{1.58} & 0.0009 \stdv{.0006} & 2.49 \stdv{1.08} & 0.0009 \stdv{.0004} \\
 & \mymethodBK       & 2.28 \stdv{1.61} & 0.0008 \stdv{.0006} & 2.18 \stdv{1.57} & 0.0008 \stdv{.0006} \\

\bottomrule
\end{tabular}}
\end{table}

%% file: tables/tab_efficiency.tex
\begin{table}[t]
\centering
\caption{Running time (in minutes) comparison of attacks.}
\label{tab:running_time}
\resizebox{0.97\linewidth}{!}{
\begin{tabular}{c | ccc | cc} 
\toprule
\textbf{Map Resolution} & PixelMatch & PixelAvg & Raster2Vec & \mymethodBU & \mymethodBK \\
\midrule
1142 $\times$ 712   & 0.10 & 0.09 & 0.10 & 1.57 & 1.30 \\
2284 $\times$ 1424  & 0.18 & 0.18 & 0.18 & 3.58 & 3.76 \\
4568 $\times$ 2848  & 0.45 & 0.47 & 0.45 & 8.21 & 8.12 \\
\bottomrule
\end{tabular}}
\end{table}

%% file: tables/tab_decimals.tex
\begin{table}[t]
\centering
\caption{Location recovery error of \mymethod at different levels of input coordinate precision.}
\label{tab:eval_decimals}
\small
\setlength{\tabcolsep}{3pt}
\resizebox{\columnwidth}{!}{
\begin{tabular}{cc|cc|cc}
\toprule
\multirow{2}{*}{\begin{tabular}{@{}c@{}}\textbf{Coordinate} \\ \textbf{Precision}\end{tabular}} & \multirow{2}{*}{\textbf{Method}}
& \multicolumn{2}{c|}{\textbf{OpenAddresses}}
& \multicolumn{2}{c}{\textbf{Synthetic}} \\
\cmidrule(lr){3-4} \cmidrule(lr){5-6}
& & \textbf{Dist. Error (m)} & \textbf{Rel. Px. Error}
& \textbf{Dist. Error (m)} & \textbf{Rel. Px. Error} \\
\midrule
\multirow{2}{*}{4 decimals}
 & \mymethodBU & 5.74 \stdv{1.92} & 0.0021 \stdv{.0007} & 5.67 \stdv{1.41} & 0.0021 \stdv{.0005} \\
 & \mymethodBK & 3.52 \stdv{1.68} & 0.0013 \stdv{.0006} & 3.46 \stdv{1.42} & 0.0013 \stdv{.0005} \\
\midrule
\multirow{2}{*}{5 decimals}
 & \mymethodBU & 2.00 \stdv{.67} & 0.0007 \stdv{.0002} & 1.87 \stdv{.54} & 0.0007 \stdv{.0002} \\
 & \mymethodBK & 1.32 \stdv{.66} & 0.0005 \stdv{.0003} & 1.24 \stdv{.55} & 0.0005 \stdv{.0002} \\
\midrule
\multirow{2}{*}{6 decimals}
 & \mymethodBU & 1.81 \stdv{.68} & 0.0007 \stdv{.0003} & 1.69 \stdv{.51} & 0.0006 \stdv{.0002} \\
 & \mymethodBK &  0.95 \stdv{.64} & 0.0003 \stdv{.0002} & 0.88 \stdv{.52} & 0.0002 \stdv{.0001} \\
\bottomrule
\end{tabular}}
\end{table}

%% file: tables/tab_boundary.tex
\begin{table}[t]
\centering
\caption{Location recovery error (meters) of \mymethod using the four-neighbor boundary set and the extended set that further includes the four-neighbors of the boundary pixels.}
\label{tab:boundary}
\small
\setlength{\tabcolsep}{3pt}
\resizebox{\columnwidth}{!}{
\begin{tabular}{cc | cc | cc}
\toprule
\multirow{2}{*}{\textbf{Background}} & \multirow{2}{*}{\textbf{Method}}
& \multicolumn{2}{c}{\textbf{OpenAddresses}}
& \multicolumn{2}{c}{\textbf{Synthetic}} \\
\cmidrule(lr){3-4} \cmidrule(lr){5-6}
&
& Four-neighbor & Extended
& Four-neighbor & Extended \\
\midrule

\multirow{2}{*}{White canvas}
 & \mymethodBU & 1.67 \stdv{.61} & 1.72 \stdv{.59} & 1.61 \stdv{.50} & 1.68 \stdv{.50} \\
 & \mymethodBK & 0.91 \stdv{.62} & 0.91 \stdv{.59} & 0.81 \stdv{.50} & 0.81 \stdv{.49} \\
\midrule

\multirow{2}{*}{Street}
 & \mymethodBU & 1.81 \stdv{.66} & 1.80 \stdv{.64} & 1.69 \stdv{.51} & 1.65 \stdv{.53} \\
 & \mymethodBK & 0.95 \stdv{.64} & 0.95 \stdv{.64} & 0.88 \stdv{.52} & 0.88 \stdv{.54} \\
\midrule

\multirow{2}{*}{Satellite}
 & \mymethodBU & 10.50 \stdv{3.64} & 10.41 \stdv{3.61} & 9.41 \stdv{3.25} & 9.41 \stdv{3.11} \\
 & \mymethodBK & 2.13 \stdv{.67} & 2.13 \stdv{.67} & 2.73 \stdv{.52} & 2.73 \stdv{.52} \\

\bottomrule
\end{tabular}
}
\end{table}

%% file: 5_4_insights.tex
\begin{figure}[t]
    \centering
    \subfigure[Box plot illustrating the distribution of recovery errors across overlapping dots. Outliers are represented as circles.]
    {
    \includegraphics[width=0.56\linewidth]{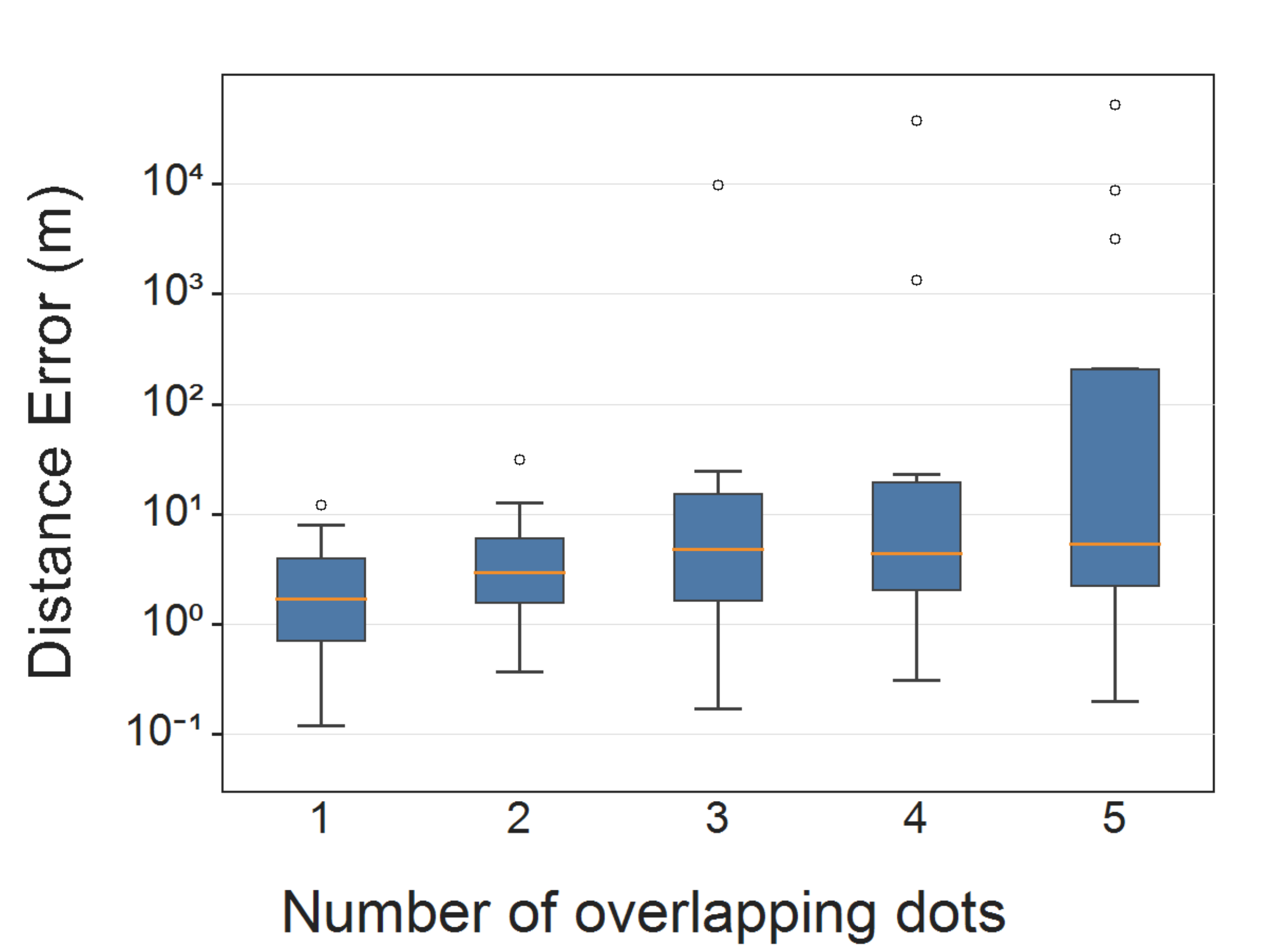}
    \label{fig:box_plot}
    }
    \hfill
    \subfigure[Top: Hard-to-attack overlapping dots. Bottom: Easy-to-attack dots.]
    {
    \includegraphics[width=0.37\linewidth]{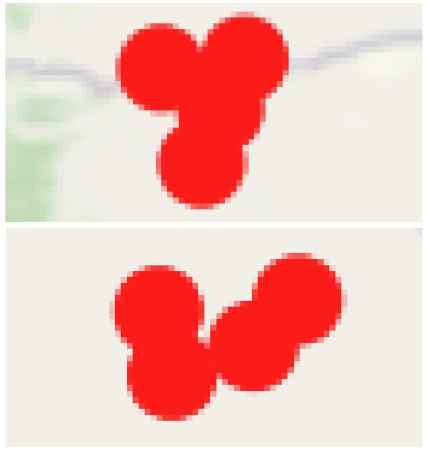}
    \label{fig:dots_demo}
    }
    \caption{(a) Analysis of recovery error distribution across varying numbers of overlapping dots. (b) Examples of easy-to-attack and hard-to-attack overlapping dots.}
\end{figure}

\subsection{In-depth Analysis (RQ3)}
\label{sec:exp_insights}

While the previous experiments mainly examined privacy risks in dot maps by considering recovery error at an aggregate level, privacy can also be viewed through a worst-case lens~\cite{dwork_dp,ccs13membership}. 
In this section, we take a finer-grained view by analyzing the variability of recovery errors across individual dots, investigating why some dots are more vulnerable to attacks while others remain resistant.

\mypara{Recovery Error Distribution}
We first plot the distribution of recovery errors for all dots on a small-scale map using \mymethodBU.
As shown in~\Cref{fig:box_plot}, we observe that dots exhibit different recovery difficulties.
While the median error is under 2 meters, some ``easy-to-attack'' instances (in the first quartile of the box plot) can be recovered with near-perfect accuracy.
Conversely, ``hard-to-attack'' instances (above the third quartile) exhibit greater resilience against our attacks.
Notably, overlapping dots, especially those with five overlaps, exhibit higher recovery errors compared to isolated dots. 
This suggests two key points: 
(i) the privacy risks associated with individual dots are not uniform across the map, 
and (ii) a large portion of dots in the map are vulnerable to our attacks because their recovery error is within a small range.
In the following analysis, we conduct case studies on both easy-to-attack and hard-to-attack dots to explore the underlying reasons for this disparity.

\myquestion{Why Some Dots are Hard to Attack}
We select a representative hard-to-attack dot with high recovery error, shown in the top part of~\Cref{fig:dots_demo}.
This dot lies in a tightly clustered region where individual anti-aliasing boundary pixels are less distinct.
Because the perceptual loss relies on boundary pixels, having only a few valid ones makes it difficult to accurately estimate the visual mismatch between the generated and target maps.
This effect is most severe for dots at the center of a cluster, where nearly all boundary pixels are shared with neighboring dots, leaving little information for precise localization.
As a result, recovery errors for such dots are significantly higher.

\myquestion{Why Some Dots are Easy to Attack}
We select a representative example from the lower quartile of the recovery error distribution, focusing on dots with four overlaps.
As shown in the bottom part of~\Cref{fig:dots_demo}, although they also form a cluster of four dots, these dots are more widely separated, preserving a more complete ring of anti-aliasing boundary pixels around each centroid.
This provides a stronger signal for the perceptual loss, enabling more accurate measurement of the visual mismatch between the generated and target maps.
Furthermore, the background beneath these dots is visually simple, with sharp contrast between the dot color and the background, making the anti-aliasing artifacts particularly prominent.
Together, these factors allow the adversary to estimate dot centroids with extremely high precision.

\input{tables/tab_background_property}

\mypara{Impact of Background Complexity and Contrast}
We further study how the local background affects attack performance.
Specifically, we describe each isolated dot by two properties of the background at its boundary pixels:
(i)~\textit{complexity}, the standard deviation of the background colors, and
(ii)~\textit{contrast}, the average color difference between these background pixels and the dot color.
We bin each property into low and high using the median as the threshold across both datasets (\ie 9.13 for complexity and 479.25 for contrast) and evaluate our methods on the street-map background.
As shown in~\Cref{tab:background_property}, both attacks remain accurate and robust across all four categories.

%% file: tables/tab_background_property.tex
\begin{table}[t]
\centering
\caption{Location recovery error (meters) of \mymethod under varying contrast and complexity of nearby background.}
\label{tab:background_property}
\small
\setlength{\tabcolsep}{3pt}
\resizebox{\columnwidth}{!}{
\begin{tabular}{cc | cc | cc}
\toprule
\multirow{2}{*}{\textbf{Complexity}} & \multirow{2}{*}{\textbf{Method}}
& \multicolumn{2}{c}{\textbf{OpenAddresses}}
& \multicolumn{2}{c}{\textbf{Synthetic}} \\
\cmidrule(lr){3-4} \cmidrule(lr){5-6}
&
& High Contrast & Low Contrast
& High Contrast & Low Contrast \\
\midrule

\multirow{2}{*}{High}
 & \mymethodBU & 1.14 \stdv{.71} & 1.97 \stdv{.90} & 1.66 \stdv{.79} & 1.73 \stdv{.70} \\
 & \mymethodBK & 0.68 \stdv{.31} & 0.79 \stdv{.47} & 0.95 \stdv{.48} & 0.89 \stdv{.35} \\
\midrule

\multirow{2}{*}{Low}
 & \mymethodBU & 2.00 \stdv{.76} & 1.59 \stdv{.70} & 1.65 \stdv{.77} & 1.72 \stdv{.73} \\
 & \mymethodBK & 1.20 \stdv{.54} & 0.98 \stdv{.40} & 0.83 \stdv{.43} & 0.94 \stdv{.27} \\

\bottomrule
\end{tabular}
}
\end{table}%

%% file: 6_defense.tex
\section{Mitigation Strategies}
\label{sec:mitigation}

Our findings suggest that even seemingly benign dot maps that cover a large geographic region can reveal concealed, high-precision location information through our attacks. 
Notably, in many maps, we were able to pinpoint a significant proportion of individuals’ locations with accuracy within 1 meter.
Thus, rigorous guidelines are needed to ensure the safe publication of dot maps.
In this section, we discuss several potential mitigation strategies and propose a risk assessment tool to evaluate the privacy risks of dot maps.

\mypara{Potential Mitigation Strategies}
We consider the following mitigation strategies to defend against location recovery attacks:
\begin{itemize}
    \item \textit{Publishing Maps Without Anti-Aliasing.}
    While anti-aliasing is enabled by default in all map visualization tools we are aware of, some tools (\eg QGIS) allow it to be disabled.
    Since \mymethod relies on anti-aliasing artifacts to reverse-engineer dot locations, disabling it could be an effective defense against our attacks.
    \item \textit{Geo-masking.}
    A significant body of work~\cite{leitner2004cartographic, stinchcomb2004procedures, cassa2006context} has developed geo-masking techniques to mitigate privacy risks in location data.
    Here, we employ a simple approach~\cite{kwan2004protection, zandbergen2014confidentiality} that adds random noise to the raw location data by displacing each point in a random direction within a fixed radius (50 or 100 meters).
    \item \textit{Location Quantization.}
    Another straightforward mitigation involves reducing the precision of the original location data (\eg from seven decimal places to three or two), shifting dots from their true positions, and thereby limiting recovery accuracy.
\end{itemize}

Note that these mitigation strategies are chosen because they are simple and widely known.
While more sophisticated methods exist~\cite{shokri2011quantifying, bindschaedler2016synthesizing, zandbergen2014confidentiality}, we focus on these approaches to illustrate the effectiveness of representative mitigation classes against our attacks.

\input{tables/tab_mitigation_1}

\mypara{Mitigation Performance}
We apply these mitigations to a small-scale map and evaluate the performance of our attacks.
\fullorcamera{The results are shown in~\Cref{tab:mitigation_bk} and~\Cref{tab:mitigation_bu}, respectively.}{The results for \mymethodBK are shown in~\Cref{tab:mitigation_bk}; the corresponding results for \mymethodBU are provided in the full version.}
All three strategies effectively degrade recovery accuracy, increasing errors from around 1 meter to tens or hundreds of meters and rendering the attack ineffective for identifying individual locations.

\mypara{Privacy Risk Assessment Tool}
While privacy regulations~\cite{hipaa,gdpr,ccpa} mandate the protection of sensitive data, they lack concrete guidelines for dot map publishing.
Motivated by this, we propose a privacy risk assessment tool to help researchers evaluate the privacy risks of their dot maps. 
We note that the success of location recovery depends on two factors: (1) the recovery accuracy of the attack and (2) the population density of the geographic region.


Based on this, we develop a privacy assessment tool that adaptively adjusts location quantization according to local population density.
The tool takes as input a set of GPS locations, a user-specified anonymity level $k$, and a publicly available population density layer; in our implementation, we use WorldPop~\cite{populationmap1}. For each location, it identifies the finest coordinate precision (\ie the largest number of decimal places) such that the corresponding spatial cell is expected to contain at least $k$ residents.
The released location is indistinguishable from those of at least $k$ residents, thereby achieving $k$-anonymity~\cite{k-anonymity}. This approach offers practitioners an intuitive and flexible way to balance privacy protection with map usability while supporting compliance with privacy regulations such as the GDPR~\cite{gdpr}, CCPA~\cite{ccpa}, and HIPAA~\cite{hipaa}.

\input{tables/tab_assessment_tool}

\mypara{Effectiveness of the Assessment Tool}
We evaluate whether the assessment tool can mitigate our attack.
Specifically, for each anonymity level $k$, we run the tool on the OpenAddresses dataset to obtain the recommended coordinate precision of every location, \ie the number of decimal places kept in its GPS latitude and longitude.
We then quantize each location to its recommended precision, render the resulting map, and attack it with \mymethodBK.
For each recovered location, we measure its anonymity set size, \ie the number of residents within the spatial cell containing the recovered coordinate.
Both metrics are averaged over all dots on the map.
As shown in~\Cref{tab:assessment_tool}, increasing $k$ leads the tool to retain fewer decimal places, from 2.44 at $k=10$ to 1.85 at $k=100$.
Because the tool assigns an integer precision to each location, three and two decimal places correspond to approximately hundred-meter- and kilometer-scale spatial precision, respectively, while one decimal place corresponds to approximately ten-kilometer-scale precision.
Thus, this quantization could potentially reduce utility for large-scale maps, while having less impact on maps intended only to show broad regional trends.
Meanwhile, the mean anonymity set size increases from 168.25 to 1,472.24 residents as $k$ increases from 10 to 100.
The mean anonymity set size exceeds the target $k$ in all settings, with the larger margins arising because many dots fall in densely populated areas.
These results demonstrate that the tool effectively mitigates our attack while making the resulting privacy--utility trade-off explicit.

%% file: tables/tab_mitigation_1.tex
\begin{table}[t]
\centering
\caption{Mitigation strategies against \mymethodBK.}
\label{tab:mitigation_bk}
\resizebox{\linewidth}{!}{%
\begin{tabular}{lcc|cc}
\toprule
\multirow{2}{*}{\textbf{Mitigation}} & \multicolumn{2}{c|}{\textbf{OpenAddresses}} & \multicolumn{2}{c}{\textbf{Synthetic}} \\
\cmidrule(lr){2-3} \cmidrule(lr){4-5}
& \textbf{Dist. Error (m)} & \textbf{Rel. Px. Error} & \textbf{Dist. Error (m)} & \textbf{Rel. Px. Error} \\
\midrule
w/o Anti-aliasing & 235.98 \stdv{166.52} & 0.0874 \stdv{.0656} & 232.31 \stdv{105.65} & 0.0860 \stdv{.0416} \\
Geo-masking (radius: 100 m) & 100.33 \stdv{.91} & 0.0372 \stdv{.0004} & 100.21 \stdv{.93} & 0.0371 \stdv{.0004} \\
Geo-masking (radius: 50 m) & 49.92 \stdv{.85} & 0.0185 \stdv{.0003} & 50.16 \stdv{.87} & 0.0186 \stdv{.0003} \\
Quantization (3 Decimals) & 36.32 \stdv{12.18} & 0.0135 \stdv{.0048} & 35.26 \stdv{12.75} & 0.0131 \stdv{.0050} \\
Quantization (2 Decimals) & 400.75 \stdv{141.60} & 0.1484 \stdv{.0558} & 393.76 \stdv{123.64} & 0.1458 \stdv{.0487} \\
\midrule
None   & 0.95 \stdv{.64} & 0.0003 \stdv{.0002} & 0.88 \stdv{.52} & 0.0002 \stdv{.0001} \\
\bottomrule
\end{tabular}}
\end{table}

%% file: tables/tab_assessment_tool.tex
\begin{table}[t]
\centering
\caption{Effectiveness of the proposed assessment tool.}
\label{tab:assessment_tool}
\small
\setlength{\tabcolsep}{3pt}
\resizebox{\columnwidth}{!}{
\begin{tabular}{c | cc}
\toprule
\textbf{Anonymity Level $k$}
& \textbf{Recommended Decimals}
& \textbf{Anonymity Set Size} \\
\midrule
10  & 2.44 \stdv{.70} & 168.25 \stdv{268.82} \\
20  & 2.15 \stdv{.63} & 556.96 \stdv{649.06} \\
50  & 1.90 \stdv{.51} & 1,204.71 \stdv{2,129.87} \\
100 & 1.85 \stdv{.47} & 1,472.24 \stdv{2,280.86} \\
\bottomrule
\end{tabular}
}
\end{table}

%% file: 0_related.tex
\section{Related Work}
\label{sec:related}

Dot maps are increasingly popular tools for visualizing the spatial distribution of individuals and events~\cite{dotmaplist, soetens2017dot, martinez1989geographic}. 
In articles and publications, dot maps are most commonly shared as raster images, reflecting the conventions of print media and the convenience of distributing fixed image formats~\cite{matiashuk2015taxonomic, dziuba2025supporting, koktava2023options}.

\mypara{Privacy Risks with Dot Maps}
Dot maps are frequently used to display sensitive personal data, such as patient locations and crime incident locations. 
For example, Armstrong~\cite{armstrong2002geographic} highlighted that in epidemiological and criminal investigations, it is common for dot maps to have a one-to-one correspondence between each dot and a specific case. 
A significant body of research~\cite{kounadi2014geoprivacy, brownstein2006unsupervised, curtis2006spatial, zandbergen2014confidentiality, leitner2007novices} demonstrates that these dots can be reverse-engineered to re-identify precise locations, posing serious privacy risks.
For instance, Brownstein et al.~\cite{brownstein2006unsupervised} found that over 26\% of locations from presentation-quality maps and over 79\% from publication maps could be accurately identified. 
Kounadi et al.~\cite{kounadi2014geoprivacy} identified 41 articles between 2005 and 2012 that disclosed over 68,000 home addresses. 
These studies raise ethical and security concerns, especially for individuals with stigmatized conditions (\eg mental illness), as they could be targeted. 

Most existing work focuses on large-scale maps that cover a limited geographic area, with little research addressing the privacy risks of small-scale dot maps that span broader regions.
Such maps have been created for regions such as Germany and the Netherlands~\cite{soetens2017dot}, Cameroon~\cite{cameroonmap2018}, and Thailand~\cite{thailandmap2025}, as listed in~\cite{dotmaplist}. 
While dot maps covering larger regions are not uncommon, their privacy implications remain largely unexplored.
Furthermore, existing geo-location privacy studies have not explored the use of anti-aliasing for location recovery. 
This is a key focus of our work, where we investigate how these map rendering techniques can be leveraged to recover high-precision location information.

\mypara{Privacy Protection Strategies for Locations and Maps}
Many studies propose geo-masking strategies to mitigate privacy risks in location-based data. 
One early approach is dot aggregation, where dot locations are aggregated at either the midpoint of the street segment or at the nearest street intersection~\cite{leitner2004cartographic, kounadi2014geoprivacy}.
Another common technique is random perturbation, which introduces random noise to location coordinates. 
Various perturbation methods have been studied, including random direction and fixed radius~\cite{kwan2004protection, zandbergen2014confidentiality}, random perturbation within a circle~\cite{armstrong1999geographically, zimmerman2008quantifying}, Gaussian displacement~\cite{zimmerman2008quantifying, cassa2008re}, donut masking~\cite{stinchcomb2004procedures, lu2012considering}, and bimodal Gaussian displacement~\cite{cassa2006context}.
Several studies extend quantitative privacy notions, such as k-anonymity~\cite{k-anonymity} and differential privacy~\cite{dwork_dp}, to geo-location data, and develop location-preserving techniques~\cite{bindschaedler2016synthesizing, shokri2011quantifying, el2009globally, wieland2008revealing}.

Deploying these defenses for dot map publications requires understanding the trade-off between privacy and map usability. 
While broad privacy regulations~\cite{hipaa,ccpa,gdpr} mandate the protection of personal locations, they lack technical specifications for visual data dissemination.
One needs to choose an appropriate defense level to satisfy privacy requirements while achieving good visualization readability.
To address this, we introduce a risk assessment framework. 
Using a population density map, researchers can select an appropriate coordinate quantization precision tailored to their specific privacy and utility requirements, offering a flexible trade-off between privacy and map usability.


\mypara{Image Vectorization and Deblurring}
The computer graphics community has studied the problem of recovering vector representations from raster images, commonly referred to as image vectorization or deblurring~\cite{kopf2011depixelizing, hoshyari2018perception, dominici2020polyfit, yang2023subpixel, pradhan2022vectorgraphics_survey}. 
These methods aim to reconstruct smooth, resolution-independent geometric shapes from pixelated inputs. 
While not designed for privacy analysis, these works share our observation that anti-aliasing encodes sub-pixel information.
However, these approaches differ from ours in several fundamental ways.
First, their objective is visual reconstruction (\eg recovering region topology and color palettes), whereas ours is the extraction of high-precision geographic coordinates. 
Second, they treat anti-aliasing as visual degradation to be eliminated in pursuit of sharp boundaries, whereas we exploit it to reverse-engineer the precise location. 
Third, advanced vectorization approaches~\cite{li2020diffvg} often require a white-box differentiable rendering pipeline, whereas our framework treats the map renderer as a black box, making it applicable to any visualization tool. 
Our empirical evaluation demonstrates that applying a standard vectorization tool (\ie Raster2Vec in QGIS) to our task yields location-recovery errors orders of magnitude larger than those of our proposed methods.

%% file: 7_discussion.tex
\section{Conclusion}
\label{sec:conclusion}

In this paper, we systematically study the privacy risks of dot maps by proposing \mymethod, an automated high-precision location recovery framework.
\mymethod is an optimization-based algorithm that exploits anti-aliasing artifacts in dot maps for precise location estimation.
Extensive experiments across different datasets and map configurations demonstrate the effectiveness and robustness of the proposed method.
We also explore several mitigation strategies and introduce a privacy assessment tool to help practitioners evaluate and mitigate the privacy risks of their dot maps.
Our work reveals a new attack vector for recovering highly precise location information from dot maps and opens new directions for analyzing the privacy risks of spatial data visualizations.

%% file: 0_open.tex
\section{Open Science}

Our artifact includes (i) source code for proposed attacks, (ii) the benchmarks, and (iii) the privacy assessment tool.
The repository is available at \url{https://github.com/PuddlesPenguin/AutoLocate/}.





%% file: 0_ethic.tex
\section{Ethics Considerations}

\ifcameraready\else
Our research investigates the privacy risks associated with dot maps, specifically focusing on high-precision location recovery from rasterized maps. 
Since dot maps are widely used to visualize sensitive data (\eg patient home addresses and crime locations), we recognize our responsibility to carefully assess the ethical implications of our findings. 
We have undertaken this assessment using the framework outlined in the Menlo Report, while adhering to the ethical guidelines set forth by CCS 2026.

\mypara{Stakeholder-Based Analysis}
This research involves several key stakeholders, each impacted by our findings in different ways:

\begin{itemize}
    \item \textit{Researchers and Map Creators.}
    Our primary audience consists of researchers and creators of dot maps. 
    We provide these practitioners with a deeper understanding of the privacy risks in dot maps, along with a concrete tool for assessing the risks of their own maps. 
    Additionally, we propose and validate mitigation strategies to address these risks.
    \item \textit{Data Subjects.}
    The data subjects in this context are the individuals whose sensitive location data is visualized on dot maps. 
    In this paper, our experiments were conducted using synthetic/public datasets, and no specific individuals or proprietary dot maps were targeted.
    Furthermore, we believe it is important to raise awareness about these underlying privacy risks and prevent potential privacy threats to individuals in the future.
    \item \textit{Map Software Developers.}
    The developers of map visualization platforms (\eg QGIS, GeoPandas, and R) are also stakeholders in this research, as our attack exploits a default rendering feature (\ie anti-aliasing) present in these platforms. 
    By publishing this work, we aim to provide developers with insights to incorporate techniques that can mitigate such privacy risks.
    \item \textit{Adversaries.}
    Our methods could be maliciously used by adversaries to identify individuals or specific locations from dot maps. 
    However, it is important to note that these risks already existed prior to our research. 
    We believe that by raising awareness of these risks, we can help mitigate broader privacy concerns. 
    Additionally, we discuss effective mitigation strategies to minimize the likelihood of malicious use of this research.
\end{itemize}
\fi

\mypara{Ethical Justification}
Dot maps are commonly used in sensitive domains,
where the potential risks of exposing individuals' locations are significant. 
Given that these risks are not always well understood, we believe it is crucial to disclose the vulnerabilities associated with publishing dot maps. 
While we recognize that malicious actors could exploit our findings, we believe that proactively sharing this knowledge enables the research community to address these privacy risks before they are exploited in real-world scenarios.
We encourage the community to use our results to develop stronger privacy protections. 

\mypara{Responsible Disclosure}
We have disclosed our findings to the developers of the map visualization platforms evaluated in this paper, including QGIS (through its security team), Matplotlib (which GeoPandas relies on for rendering), and the maintainer of the R maps package.
We have also reported our findings to the U.S. CDC, which publishes cartographic guidelines~\cite{cdc2012cartographic} for sensitive data.
Several of these stakeholders have engaged with us, and we are working with them to address this risk.
 \cameraonly{A full discussion of the ethical considerations is provided in the full version.}

%% file: 8_0_appendix.tex
\input{8_1_appendix_survey}

\input{8_2_appendix_implementation}

%% file: 8_1_appendix_survey.tex
\section{Literature Survey of Dot Map Usage}
\label{appendix:survey}

To demonstrate the widespread prevalence of dot maps, we conducted a broad survey covering both academic research and public media. 
Our methodology utilized academic databases (Google Scholar, Semantic Scholar) for research publications and general search engines for journalistic and governmental examples. 
We combined general terms for the visualization technique (\eg ``dot map'', ``point map'', ``spatial distribution'') with domain-specific keywords. 
The queries for each category were structured as follows: 
\begin{itemize}
    \item For \textbf{Public Health}, we used (``public health'' OR ``epidemiology'') AND ``dot map'' AND (``patient location'' OR ``case distribution'').
    \item For \textbf{Criminology}, we used ``crime mapping'' AND ``point map'' AND ``incident location''.
    \item For \textbf{Ecology}, we used ``ecology'' AND ``point map'' AND (``endangered species'' OR ``presence-only data'').
    \item For \textbf{Social Science}, we used (``social science'' OR ``demography'') AND ``dot density map'' AND (``household demographics'' OR ``racial dot map'').
    \item For \textbf{Education}, we used ``education'' AND ``spatial analysis'' AND ``student residence''.
    \item For \textbf{Archaeology}, we used ``archaeology'' AND ``point map'' AND ``artifact find spot''.
\end{itemize}

For every result, we manually inspected the source (whether an academic paper, news article, or web report) to confirm it utilized a raster dot map where each dot represents a discrete data point (\eg one person or incident). 
If a source was not a direct match, we examined its citations or references to identify related examples. 
The resulting selected publications are shown in~\Cref{tab:survey}. 
Note that this is an exemplary list intended to showcase the prevalence of dot map usage, not a comprehensive or systematic review. 
We refer the reader to~\cite{dotmaplist} for a more complete survey of spatial visualization usage in academic research.

\begin{figure}[t]
\centering
\subfigure[Excerpted from Figure 1 of~\cite{foraker2022enabling}.]
{
    \includegraphics[width=0.39\linewidth]{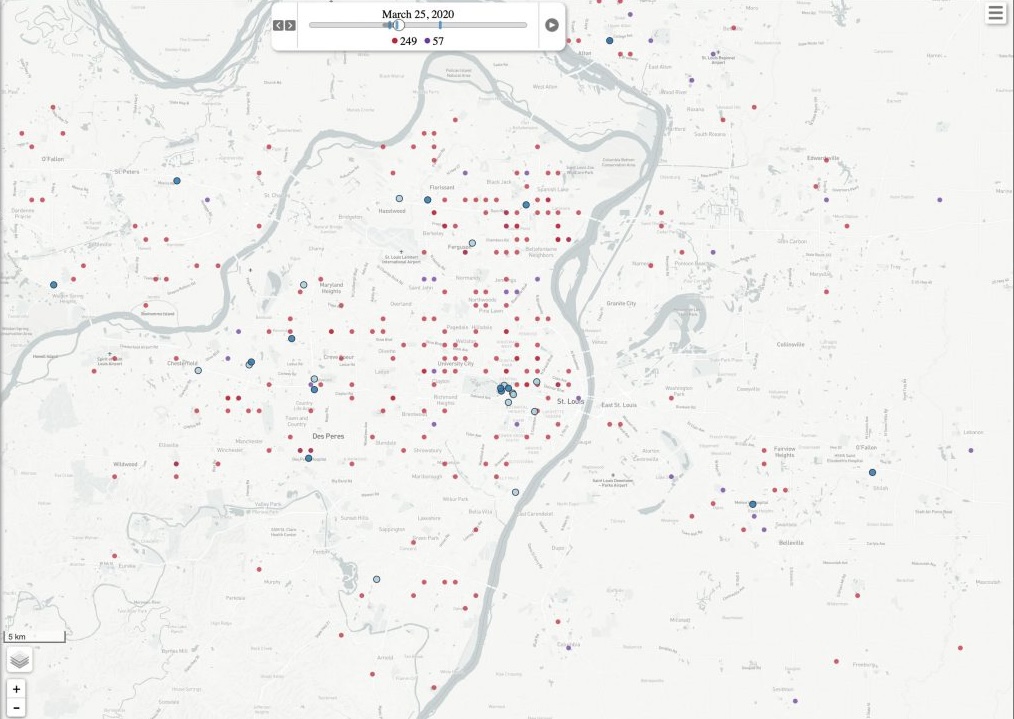}
    \label{fig:example_1}
}
\hfill
\subfigure[Excerpted from Figure 1 of~\cite{obaldia2015panama}.]
{
    \includegraphics[width=0.5\linewidth]{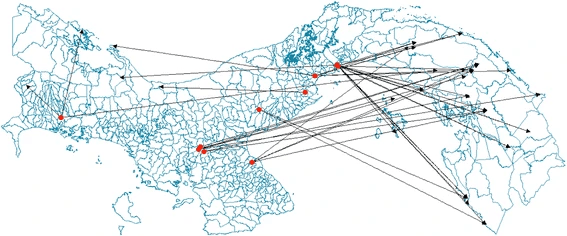}
    \label{fig:example_2}
}
\caption{Examples of dot maps in published papers, where most dots are isolated.}
\label{fig:published_maps}
\end{figure}

\mypara{Isolated Dots in Published Maps}
We examine real published dot maps to assess whether our attack is applicable in practice.
Specifically, we manually count the dots in two representative maps from~\Cref{tab:survey}, treating a dot as isolated if it does not overlap with any other dot.
As shown in~\Cref{fig:published_maps}, isolated dots account for the majority of dots in both maps.
The map from~\cite{foraker2022enabling} contains 306 dots, of which 287 (93.8\%) are isolated, while 6 out of the 9 (red) dots (66.7\%) in a map from~\cite{obaldia2015panama} are isolated.
\mymethod achieves strong location recovery performance on isolated dots; their prevalence in published maps indicates that our attack is applicable in practice.

%% file: 8_2_appendix_implementation.tex
\section{Implementation Details}
\label{appendix:implementation}

\input{tables/tab_implementation_comparison}

\input{tables/tab_map_config}

\input{tables/tab_mitigation_2}

\input{tables/alg_boundary}

\mypara{Implementations of \mymethod}
By default, \mymethod renders one candidate map for each search direction, in which all dots are shifted simultaneously, which we refer to as the \textit{batched implementation}.
This design keeps the number of rendered maps in each optimization iteration independent of the number of dots.
An alternative is to render a separate candidate map for each dot and compute its loss individually, which we refer to as the \textit{per-dot implementation}.
We compare the two implementations in terms of recovery error and runtime.
As shown in~\Cref{tab:implementation_comparison}, the two implementations achieve comparable median recovery errors for both isolated and overlapping dots.
The per-dot implementation yields only modest improvements in median error in a few cases (\eg from 27.00\,m to 22.00\,m for four overlapping dots on OpenAddresses using \mymethodBU).
In contrast, it requires over an order of magnitude more runtime.
We therefore adopt the batched implementation for its efficiency.

\mypara{Boundary Pixel Identification}
\Cref{alg:boundary} presents the procedure for identifying the boundary pixels of each dot.
It takes as input the dot regions $\mathcal{P}$ and the estimated dot locations $\mathcal{C}$ obtained in Phase~1 of~\Cref{alg:framework}, and returns the boundary pixel set $\mathcal{S}_i$ for each dot $\mathbf{c}_i$.
For each dot region, the algorithm examines the four neighboring pixels of every pixel in the region and collects those that fall outside the region.
Each collected pixel is then assigned to its nearest dot, forming the boundary pixel set for that dot.

%% file: tables/tab_implementation_comparison.tex
\begin{table*}[t]
\centering
\caption{Performance comparison of the batched and per-dot implementations of \mymethod.}
\label{tab:implementation_comparison}
\small
\setlength{\tabcolsep}{3pt}
\resizebox{1.85\columnwidth}{!}{
\begin{tabular}{ccc | ccccc | c}
\toprule
\multirow{2}{*}{\textbf{Dataset}} & \multirow{2}{*}{\textbf{Method}} & \multirow{2}{*}{\textbf{Implementation}}
& \multicolumn{5}{c|}{\textbf{Dist. Error (m) by \# Overlapping Dots}}
& \multirow{2}{*}{\textbf{Runtime (min)}} \\
\cmidrule(lr){4-8}
& & & 1 & 2 & 3 & 4 & 5 & \\
\midrule

\multirow{4}{*}{OpenAddresses}
 & \multirow{2}{*}{\mymethodBU} & Batched & 1.81 \stdv{.66} & 7.34 \stdv{102.29} & 8.89 \stdv{185.73} & 27.00 \stdv{1046.78} & 16.12 \stdv{11828.85} & 3.58 \\
 &                              & Per-dot & 1.42 \stdv{.57} & 7.41 \stdv{111.63} & 8.79 \stdv{162.48} & 22.00 \stdv{987.31} & 15.70 \stdv{12417.52} & 87.7 \\
 \cmidrule(lr){2-9}
 & \multirow{2}{*}{\mymethodBK} & Batched & 0.95 \stdv{.64} & 2.26 \stdv{1.04} & 4.31 \stdv{32.54} & 4.86 \stdv{241.87} & 5.46 \stdv{10264.44} & 3.76 \\
 &                              & Per-dot & 0.85 \stdv{.68} & 2.34 \stdv{.95} & 3.31 \stdv{35.82} & 4.74 \stdv{207.41} & 4.95 \stdv{11203.76} & 73.3 \\
\midrule

\multirow{4}{*}{Synthetic}
 & \multirow{2}{*}{\mymethodBU} & Batched & 1.69 \stdv{.51} & 3.00 \stdv{1.66} & 4.79 \stdv{50.39} & 4.39 \stdv{161.92} & 5.37 \stdv{641.73} & 3.58 \\
 &                              & Per-dot & 1.61 \stdv{.46} & 3.02 \stdv{1.81} & 4.56 \stdv{43.27} & 4.52 \stdv{178.64} & 5.66 \stdv{587.19} & 137.3 \\
 \cmidrule(lr){2-9}
 & \multirow{2}{*}{\mymethodBK} & Batched & 0.88 \stdv{.52} & 1.82 \stdv{.70} & 2.17 \stdv{.99} & 2.50 \stdv{15.54} & 2.92 \stdv{1158.51} & 3.76 \\
 &                              & Per-dot & 0.76 \stdv{.55} & 1.86 \stdv{.64} & 1.95 \stdv{1.09} & 2.53 \stdv{13.91} & 3.83 \stdv{1274.83} & 145.1 \\

\bottomrule
\end{tabular}}
\end{table*}

%% file: tables/tab_map_config.tex
\begin{table}[t]
    \centering
    \caption{Map configuration space. The default setting for each dimension is marked in bold.}
    \label{tab:map_config}
    \resizebox{0.95\linewidth}{!}{
    \begin{tabular}{llc}
        \toprule
        \textbf{Dimension} & \textbf{Values} & \textbf{\# Settings} \\
        \midrule
        Map scale       & \textbf{Small (US)}, Medium (OH), Large (Austin, TX) & 3 \\
        Background  & White canvas, \textbf{Street map}, Satellite imagery   & 3 \\
        Resolution  & 96, \textbf{192}, 384 DPI                              & 3 \\
        Dot geometry & \textbf{Circle}, Pentagon, Triangle                      & 3 \\
        Dot size    & 1\,mm, \textbf{2\,mm}, 3\,mm                           & 3 \\
        Format      & \textbf{PNG}, JPEG, TIFF                               & 3 \\
        Platform    & \textbf{GeoPandas}, QGIS, R                            & 3 \\
        \bottomrule
    \end{tabular}}
\end{table}

%% file: tables/tab_mitigation_2.tex
\begin{table}[t]
\centering
\caption{Mitigation strategies against \mymethodBU.}
\label{tab:mitigation_bu}
\resizebox{\linewidth}{!}{%
\begin{tabular}{lcc|cc}
\toprule
\multirow{2}{*}{\textbf{Mitigation}} & \multicolumn{2}{c|}{\textbf{OpenAddresses}} & \multicolumn{2}{c}{\textbf{Synthetic}} \\
\cmidrule(lr){2-3} \cmidrule(lr){4-5}
& \textbf{Dist. Error (m)} & \textbf{Rel. Px. Error} & \textbf{Dist. Error (m)} & \textbf{Rel. Px. Error} \\
\midrule
w/o Anti-aliasing & 235.98 \stdv{181.42} & 0.0874 \stdv{.0715} & 232.31 \stdv{120.83} & 0.0860 \stdv{.0476} \\
Geo-masking (radius: 100 m) & 100.97 \stdv{.97} & 0.0374 \stdv{.0004} & 102.15 \stdv{.87} & 0.0378 \stdv{.0003} \\
Geo-masking (radius: 50 m) & 50.72 \stdv{.85} & 0.0188 \stdv{.0003} & 50.34 \stdv{.86} & 0.0186 \stdv{.0003} \\
Quantization (3 Decimals) & 38.27 \stdv{12.22} & 0.0142 \stdv{.0048} & 36.64 \stdv{12.76} & 0.0136 \stdv{.0050} \\
Quantization (2 Decimals) & 407.37 \stdv{141.01} & 0.1509 \stdv{.0555} & 396.97 \stdv{123.69} & 0.1470 \stdv{.0487} \\
\midrule
None & 1.81 \stdv{.66} & 0.0007 \stdv{.0003} & 1.69 \stdv{.51} & 0.0006 \stdv{.0002} \\
\bottomrule
\end{tabular}}
\end{table}

%% file: tables/alg_boundary.tex
\begin{algorithm}[t]
\caption{\textbf{FindBoundaryPixels.} It identifies the boundary pixels of each dot from the target dot map.}
\label{alg:boundary}
\begin{algorithmic}[1]
\REQUIRE Dot regions $\mathcal{P}$, estimated dot locations $\mathcal{C} = \{ \mathbf{c}_1, \dots, \mathbf{c}_n \}$
\STATE $\mathcal{S}_i \gets \emptyset$ for $i = 1, \dots, n$
\FOR{each component $P \in \mathcal{P}$}
    \FOR{each pixel $(x, y) \in P$}
        \FOR{each four-neighbor $(u, v)$ of $(x, y)$}
            \IF{$(u, v) \notin P$}
                \STATE $i \gets \arg\min_{j} \lVert (u, v) - \mathbf{c}_j \rVert$
                \algcomment{assign to the nearest dot}
                \STATE $\mathcal{S}_i \gets \mathcal{S}_i \cup \{ (u, v) \}$
            \ENDIF
        \ENDFOR
    \ENDFOR
\ENDFOR
\RETURN $\{ \mathcal{S}_i \}_{i=1}^{n}$
\end{algorithmic}
\end{algorithm}